\documentclass[reqno,12pt]{amsart}

\usepackage[T1]{fontenc}
\usepackage{lmodern}

\usepackage[numbers,sort&compress]{natbib}
\usepackage{graphicx}
\usepackage{hyperref}
\usepackage{amsmath,amssymb,amsthm}
\usepackage{amsaddr}
\usepackage{mathtools}
\usepackage{textcomp}
\usepackage{algorithm}
\usepackage{algpseudocode}
\usepackage{array}
\usepackage[labelfont={rm}, justification=centering]{subfig}
\usepackage{booktabs}

\calclayout

\DeclareMathAlphabet{\mathbsl}{OT1}{cmr}{bx}{sl}  

\newcommand{\M}[1]{\mathbsl{#1}}
\newcommand{\MS}[1]{\boldsymbol{#1}}

\newcommand{\NM}[1]{\underline{#1}}

\newcommand{\D}    {\mathrm{d}}
\newcommand{\sgn}  {\operatorname{sgn}}

\newcommand{\transpose}[1]{#1^{\textsc{t}}}

\newcommand{\jump}[1]%
  {[\hspace{-0.15em}[\hspace{0.05em}#1\hspace{0.05em}]\hspace{-0.15em}]}

\newcommand{\average}[1]%
  {\{\hspace{0.05em}#1\hspace{0.05em}\}}

\newcommand{\sfrac}[2]{#1\hspace{-0.05em}/\hspace{-0.05em}#2}

\newcommand{\nel}    [1][] {N_{\textsc{e}#1}}
\newcommand{\ns}     [1][] {N_{\textsc{s}#1}}
\newcommand{\no}     [1][] {N_{\textsc{o}#1}}
\newcommand{\deltaO} [1][] {\delta_{#1\textsc{o}}}

\newcommand{\np}  {N_{\textsc{P}}}

\newcommand{\nsd} {N_{\textsc{d}}}

\newcommand{\co}  {C_{\textsc{o}}}
\newcommand{\cs}  {C_{\textsc{s}}}
\newcommand{\cd}  {C_{\textsc{d}}}
\newcommand{\md}  {M_{\textsc{d}}}

\newcommand{\IOP}{\NM{\mathcal{I}}}

\newcommand{\tsub}[1]{$_{#1}$}

\newcommand{\pbox}[1]{\makebox[2.75em][c]{$P\!=\hspace{-1.pt}#1$}}

\newcommand{\half}  {\textonehalf}

\theoremstyle{remark}
\newtheorem*{remark}{Remark}

\begin{document}

\title%
[
Robust multigrid for high-order DG: Fast Cartesian Poisson solver
]
{
Robust multigrid for high-order discontinuous Galerkin methods:
A fast Poisson solver suitable for high-aspect ratio Cartesian grids
}%

\author{J\"org Stiller}
\address{%
  Technische Universit\"at Dresden, Institute of Fluid Mechanics and
  Center for Advancing Electronics Dresden, 01062 Dresden, Germany}
\email{joerg.stiller@tu-dresden.de}

\thanks{The final article is available at \url{http://dx.doi.org/10.1016/j.jcp.2016.09.041}}

\begin{abstract}
We present a  polynomial multigrid method for nodal interior penalty and local discontinuous Galerkin formulations of the Poisson equation on Cartesian grids.
For smoothing we propose two classes of overlapping Schwarz methods.
The first class comprises element-centered and the second face-centered methods.
Within both classes we identify methods that achieve superior convergence rates,
prove robust with respect to the mesh spacing and the polynomial order, at least up to ${P=32}$.
Consequent structure exploitation yields a computational complexity of $O(PN)$, where $N$ is the number of unknowns.
Further we demonstrate the suitability of the face-centered method for element aspect ratios up to 32.
\end{abstract}


\maketitle


\section{Introduction}
\label{sec:intro}

High-order discretization methods are exciting because of their promise to deliver higher accuracy at lower cost than first and second order methods.
Much confidence has been put in the discontinuous Galerkin (DG) method because it combines multiple desirable properties of finite element and finite volume methods, including geometric flexibility, variable approximation order, straightforward adaptivity and suitability for conservation laws \cite{Cockburn2000,Hesthaven_NDGM_2008}.
Traditionally, DG methods have been used in the numerical solution of hyperbolic and convection-dominated problems.
Nevertheless, the need for implicit diffusion schemes and application to other problem classes, such as incompressible flow and elasticity, led to a growing interest in DG methods and related solution techniques for elliptic equations \cite{Arnold2001,Riviere2008}.

The most efficient elliptic solvers are based on multigrid (MG) techniques and can be classified into
polynomial or $p$-MG \cite{HAM03,FOLD05,HA06},
geometric or $h$-MG \cite{GK03,Kan04,DLV06,Kan08,KT08,SMB09} or, combining both concepts,
$hp$-MG \cite{VR12,ASV15},
and
algebraic MG \cite{PLH09,OS11,BBS12,STGSC14}.
Apart from their different coarsening strategy, polynomial and geometric multigrid are closely related to each other and can be applied with the same smoothing methods.
Early work on \mbox{$p$-MG} goes back to Helenbrook and coworkers \cite{HAM03,HA06} who explored various smoothers for DG formulations of the Poisson equation.
For isotropic grids they identified block Gauss-Seidel as the best choice, whereas more expensive line smoothing proved necessary on high-aspect-ratio grids.
At about the same time, \citet{GK03} developed $h$-MG preconditioners for Poisson and convection-diffusion problems, which also use element-based block-Gauss-Seidel methods for smoothing.
\citet{Kan04,Kan08} extended this approach to locally refined Cartesian grids in two and three space dimensions.
In both cases, polynomial and geometric multigrid, block-Gauss-Seidel smoothing yields acceptable convergence rates for low to moderate polynomial degrees, e.g.\ ${\rho \approx 0.5}$ with one pre-smoothing for {$P=4$}.
However, the convergence degrades with increasing $P$, which renders the approach unfeasible for higher polynomial degrees.

Several researchers proposed algebraic multigrid methods for various DG formulations of elliptic equations.
\citet{OS11} presented a preconditioned conjugate gradient (PCG) method based on smoothed aggregation.
Using block relaxation combined with energy-minimizing prolongation it attains mesh independent convergence rates corresponding to residual reductions of about 0.3 per smoothing step for ${P \le 6}$, but degrades with increasing approximation order.
\citet{BBS12} proposed a non-smoothed aggregation approach.
For smoothing they use block relaxations which operate on extended aggregates and, hence, can be regarded as overlapping Schwarz methods.
This approach yields by far the most efficient DG-MG method reported until now.
The method proved robust with respect to the polynomial order, up to at least ${P=6}$, though the iteration count exhibits a logarithmic dependence on the mesh spacing.
For the DG method of Oden, Babu\v{s}ka and Baumann it achieved convergence rates of ${\rho \approx 0.04}$ with one pre- and post-smoothing, which corresponds to a residual reduction by a factor of 25 in one step.
The approach was also shown to work with symmetric and non-symmetric interior penalty methods, although it required nearly twice as many iterations in the latter case.
A possible drawback is the rise of cost with increasing polynomial order.
The authors did not specify the complexity of their algorithm, however the solver runtimes indicate that the cost per unknown grows as $P^3$ in 2D and $P^5$ in 3D.

To the best of our knowledge, none of the proposed MG methods is robust with respect to both, the polynomial order and the mesh spacing.
Computational complexity and robustness against high aspect ratios are further issues that need to be considered to strengthen the competitiveness of DG methods for elliptic equations.
As a step into that direction we present a new $p$-multigrid method for interior penalty and local discontinuous Galerkin discretizations of the Poisson equation on Cartesian grids.
Our approach is motivated and strongly influenced by previous work dedicated to the continuous spectral element method \cite{LF05,HSN13,HSF2016jcp,Stiller2016jsc}.
We propose two classes of multiplicative and weighted additive Schwarz methods, which use an adjustable overlap depending on the polynomial level.
The first class comprises element-centered and the second face-centered methods.
Within both classes we identify methods that
achieve superior convergence rates,
prove robust with respect to the mesh spacing and the polynomial order and
reach a computational complexity of $O(PN)$, where $N$ is the number of unknowns.
Further we demonstrate the suitability of the face-centered method for high element aspect ratios.

The paper is organized as follows:
In the next section we derive a unified nodal DG formulation of the Poisson problem comprising  the symmetric interior penalty method and the local discontinuous Galerkin method.
Then we describe the solution methods, i.e.\ Schwarz, multigrid, and inexact PCG, in Section~\ref{sec:methods}.
Section~\ref{sec:results} presents the numerical experiments and
Section~\ref{sec:conclusions} concludes the paper.


\section{Discontinuous Galerkin method}
\label{sec:dgm}


\subsection{Problem definition}
As a model problem we consider the Poisson equation
\begin{equation}
\label{eq:poisson}
-\nabla^2 u = f
\end{equation}
in the rectangular periodic domain \mbox{$\Omega=[0,l_1]\times[0,l_2]$}.
By introducing the flux vector \mbox{$\MS{\sigma} = \nabla u$}, the problem can be rewritten into the first-order system
\begin{subequations}
\label{eq:flux form}
\begin{gather}
\label{eq:flux form:1}
\MS{\sigma} = \nabla u         \, , \\
\label{eq:flux form:2}
-\nabla\cdot\MS{\sigma} = f   \, .
\end{gather}
\end{subequations}
This form serves as the starting point for the discontinuous Galerkin method.


\subsection{Spatial discretization}

The domain $\Omega$ is decomposed into \mbox{$\nel = \nel[,1] \times \nel[,2]$}
rectangular elements
\begin{equation*}
\Omega^{m_1,m_2}
= \big(x_1^{m_1 - \sfrac{1}{2}}, x_1^{m_1 + \sfrac{1}{2}}\big) \times
  \big(x_2^{m_2 - \sfrac{1}{2}}, x_2^{m_2 + \sfrac{1}{2}}\big)
\end{equation*}
with dimensions
\mbox{$\Delta x_d^{m_d} = x_d^{m_d + \sfrac{1}{2}} - x_d^{m_d - \sfrac{1}{2}}$}
for \mbox{$d = 1, 2$}.
Each element is mapped to the standard region \mbox{$(-1,1)^2$} by
\begin{gather*}
\xi_d^{m_d}(x_d) = \frac{2}{\Delta x_d^{m_d}}(x_d - x_d^{m_d - \sfrac{1}{2}}) - 1
\, .
\end{gather*}
For simplicity we use the array notations
\mbox{$\M m = (m_1, m_2)$},
\mbox{$\M x = (x_1, x_2)$} and
\mbox{$\MS \xi = (\xi_1, \xi_2)$}
where possible.

Let $\{\varphi_i(\xi)\}_{i=0}^P$ be a polynomial basis of degree $P$ in the interval $[-1,1]$.
Using a tensor-product ansatz the solution to \eqref{eq:flux form}
can be approximated in $\Omega^{\M m}$ as
\begin{gather}
\label{eq:u_h}
u_h(\M x)|_{\Omega^{\M m}}
= u^{\M m}(\MS\xi^{\M m}(\M x))
= \sum_{i,j=0}^P u_{ij}^{\M m} \varphi_i(\xi_1) \varphi_j(\xi_2)
\\ \intertext{and}
\label{eq:sigma_h}
\MS\sigma_h(\M x)|_{\Omega^{\M m}}
= \MS\sigma^{\M m}(\MS\xi^{\M m}(\M x))
= \sum_{i,j=0}^P \MS\sigma_{ij}^{\M m} \varphi_i(\xi_1) \varphi_j(\xi_2) \, .
\end{gather}
The global solution $u_h$ and the fluxes $\MS \sigma_h$ belong to the function spaces
\begin{align*}
V_h &=
  \left\{
    v \in L^2(\Omega) : v|_{\Omega^{\M m}} \in Q_p(\Omega^{\M m}) \quad
    \forall \Omega^{\M m} \subset \Omega
  \right\}
\quad\text{and}
\\
\MS\varSigma_h &=
  \left\{
    \MS\tau \in [L^2(\Omega)]^2 :
    \MS\tau|_{\Omega^{\M m}} \in [Q_p(\Omega^{\M m})]^2 \quad
    \forall \Omega^{\M m} \subset \Omega
  \right\}
\, ,
\end{align*}
respectively, where $Q_p(\Omega^{\M m})$ is the space spanned by the tensor-product polynomials $x_1^k x_2^l$ with $0 \le k,l \le P$ in $\Omega^{\M m}$.


\subsection{Elemental DG formulation}

Following \citet{Cockburn_LDG_1998} we consider discontinuous Galerkin formulations of the form:
Find
 \mbox{$u_h \in V_h$} and
 \mbox{$\MS{\sigma}_h \in \MS\varSigma_h$}
such that for all \mbox{$\Omega^{\M m} \subset \Omega$}
\begin{subequations}
\label{eq:dg:elemental}
\begin{align}
\label{eq:dg:elemental:1}
\int_{\Omega^{\M m}} \!\! \MS\tau \cdot \MS\sigma_h \,\D\Omega
= -\int_{\Omega^{\M m}} \!\! (\nabla \cdot \MS\tau) u_h \,\D\Omega
+  \int_{\partial\Omega^{\M m}} \!\! \MS\tau \cdot \hat u \M n \,\D\Gamma
\quad &
\forall \MS\tau \in [Q_p(\Omega^{\M m})]^2,
\\
\label{eq:dg:elemental:2}
\int_{\Omega^{\M m}} \!\! \nabla v \cdot \MS{\sigma}_h \,\D\Omega
= \int_{\Omega^{\M m}} \!\! v f \,\D\Omega
+ \int_{\partial\Omega^{\M m}} \!\! v \hat{\MS\sigma} \cdot \M n \,\D\Gamma
\quad &
\forall v \in Q_p(\Omega^{\M m})\, ,
\end{align}
\end{subequations}
where the numerical fluxes $\hat u$ and $\hat{\MS\sigma}$ are approximations to $u$ and to \mbox{$\MS\sigma = \nabla u$}, respectively, on the element boundary $\partial\Omega^{\M m}$.
To concretize the numerical fluxes we need some additional notation.
For given functions
\mbox{$v \in V_h$},
\mbox{$\MS\tau \in \MS\varSigma_h$} let
$v^-$, $\MS\tau^{\,-}$ and
$v^+$, $\MS\tau^{\,+}$ denote the interior and exterior traces on
$\partial\Omega^{\M m}$, respectively.
Now we define the average and jump operators by
\begin{subequations}
\label{eq:trace operators}
\begin{alignat}{2}
\average{v}
&= \frac{1}{2}\left( v^- + v^+ \right),
\qquad
&\average{\MS\tau}
&= \frac{1}{2}\left( \MS\tau^{\,-} + \MS\tau^{\,+} \right),
\\
\jump{v}
&= \left( v^- - v^+ \right) \M n\,,
\qquad
&\jump{\MS\tau}
&= \left( \MS\tau^{\,-} - \MS\tau^{\,+} \right) \cdot \M n \, .
\end{alignat}
\end{subequations}
We remark that, despite element oriented notation, these definitions are in fact element independent and equivalent to those given in \cite{Arnold2001}.

For constructing the numerical fluxes we consider the interior penalty method (IP) and the local discontinuous Galerkin method (LDG).
We closely follow the notation used in \cite{Arnold2001}.
With IP the numerical fluxes take the form
\begin{subequations}
\begin{align}
\label{flux:num:u:IP}
\hat u_{\textsc{ip}}
&= \average{u_h}
\, ,
\\
\label{flux:num:sigma:IP}
\hat{\MS\sigma}_{\textsc{ip}}
&= \average{\nabla u_h} - \mu_{\textsc{ip}} \jump{u_h}
\end{align}
\end{subequations}
and with LDG
\begin{subequations}
\begin{align}
\label{flux:num:u:LDG}
\hat u_{\textsc{ldg}}
&= \average{u_h} - \MS\beta \cdot \jump{u_h}
\, ,
\\
\label{flux:num:sigma:LDG}
\hat{\MS\sigma}_{\textsc{ldg}}
&= \average{\MS\sigma_h} + \MS\beta\jump{\MS\sigma_h} - \mu_{\textsc{ldg}} \jump{u_h}
\, .
\end{align}
\end{subequations}
Here $\mu_{\textsc{ip}}$ and $\mu_{\textsc{ldg}}$ are positive penalty functions that are defined on the edges and typically piecewise constant.
According to \cite{Arnold2001}, \mbox{$\mu_{\textsc{ldg}} > 0$} is sufficient for stability with LDG, whereas no general stability threshold is known for IP.
The auxiliary parameter $\MS\beta$ can be used to improve the sparsity of the stiffness matrix \cite{Cockburn_LDG_1998}.
Moreover, \citet{Cockburn_LDG_cartesian_2002} devised a special form of $\MS\beta$ which yields superconvergence in the $L^2$ norm when used with Cartesian grids.
This form includes constant vectors with components satisfying \mbox{$|\beta_d| = \sfrac{1}{2}$}.
As a possible drawback, however, any non-trivial choice of \mbox{$\MS\beta$} also implies a directional bias and thus breaks the symmetry of diffusive transport.


\subsection{Discrete equations}

In the following we constrain ourselves to nodal bases formed by the Lagrange polynomials to the Gauss-Lobatto-Legendre (GLL) quadrature points $\{\eta_i\}$
in the standard interval \mbox{$[-1,1]$}, see e.g. \cite{DFM02,KS05,Hesthaven_NDGM_2008}.
The GLL quadrature is used for evaluating integrals over element domains, which yields a diagonal mass matrix without degrading the overall accuracy of the method.

Before stating the discrete equations, let us introduce some notation:
${\NM u^{\M m} = [ u_{ij}^{\M m} ]}$ represents the vector of solution coefficients
in $\Omega^{\M m}$ and
${\NM u = [ \NM u^{\M m} ]}$ the global solution vector.
Underscores indicate nodal vectors and operator matrices,
and double indices refer to the directions of the tensor-product decomposition introduced in \eqref{eq:u_h} and \eqref{eq:sigma_h}.
Due to the latter, we have to distinguish between one-dimensional and two-dimensional operators.
In particular,
the Kronecker delta $\delta_{i,k}$ denotes the components of the 1D unit matrix,
${M_{i,k}^s} = \int\!\varphi_i \varphi_j\D\xi = \rho_i \delta_{i,k}$ the (approximate) standard 1D mass matrix resulting from GLL quadrature with weights $\rho_i$, and
${D_{i,k}^s = \varphi_k'(\eta_i)}$ the standard 1D differentiation matrix with respect to the GLL points \cite{DFM02,KS05}.
The 2D element mass matrix $\NM M^{\M m}$ constitutes the tensor product of the one-dimensional mass matrices $\NM M^{m_1}_1$ and  $\NM M^{m_2}_2$, i.e.,
\begin{equation}
\label{eq:element mass matrix}
M_{ij,kl}^{\M m} = M_{1,i,k}^{m_1} \, M_{2,j,l}^{m_2}
\end{equation}
with
\begin{equation}
\label{eq:1D mass matrix}
M_{d,i,k}^{m}
= \int_{x_d^{m-\sfrac{1}{2}}}^{x_d^{m+\sfrac{1}{2}}} \!
    \varphi_i(\xi_d)\varphi_k(\xi_d)
    \,\D x_d
= \frac{\Delta x_d^{m}}{2} M_{i,k}^s
\, , \quad
d = 1, 2
\, .
\end{equation}
Note that the mass matrices are diagonal, which is a welcome side effect of the GLL quadrature.

\begin{figure}
\centering
\includegraphics[scale=0.25]{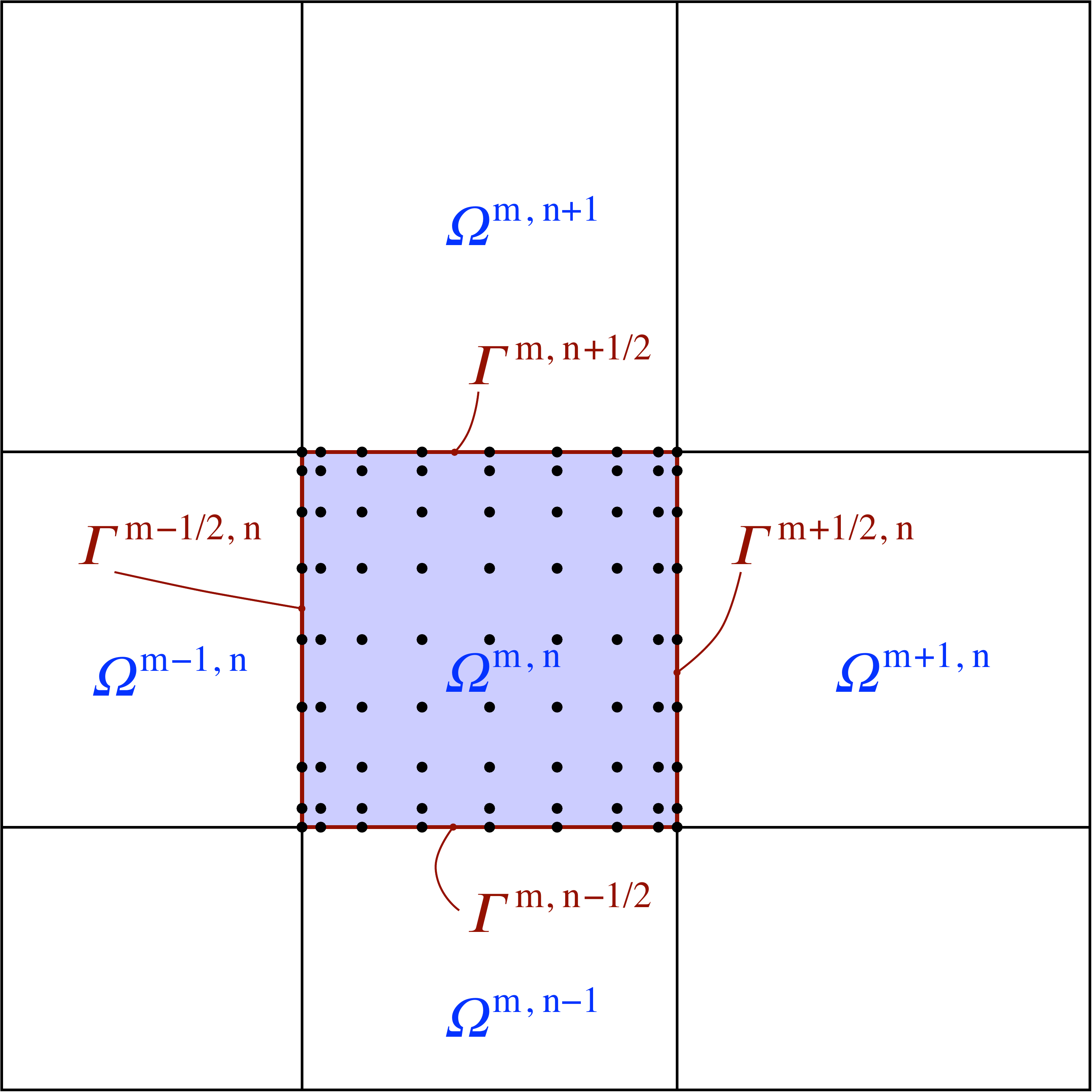}
\caption{Element domain with collocation points of order $P=8$.
  \label{fig:element}}
\end{figure}

Next we derive explicit expressions for the discrete fluxes.
Application of the divergence theorem to \eqref{eq:dg:elemental:1} yields the equivalent form
\begin{equation*}
\int_{\Omega^{\M m}} \!\! \MS\tau \cdot \MS\sigma_h \,\D\Omega
= \int_{\Omega^{\M m}} \!\! \MS\tau \cdot \nabla u_h \,\D\Omega
+  \int_{\partial\Omega^{\M m}} \!\! \MS\tau \cdot (\hat u - u_h^-) \M n \,\D\Gamma
\, .
\end{equation*}
Substituting the numerical flux \eqref{flux:num:u:IP} or \eqref{flux:num:u:LDG} for $\hat u$, respectively,
and using \eqref{eq:trace operators} we obtain
\begin{equation*}
\int_{\Omega^{\M m}} \!\! \MS\tau \cdot \MS\sigma_h \,\D\Omega
= \int_{\Omega^{\M m}} \!\! \MS\tau \cdot \nabla u_h \,\D\Omega
- \int_{\partial\Omega^{\M m}} \!\!
    \left( \frac{1}{2} + \MS\beta \cdot \M n \right) \MS\tau \cdot \jump{u_h} \,\D\Gamma
\, ,
\end{equation*}
where \mbox{$\MS\beta$} vanishes for IP and is constant for LDG.
Choosing
${\MS\tau|_{\Omega^{\M m}} = \varphi_i(\xi_1) \varphi_j(\xi_2) \M e_d}$, substituting the approximate solution (\ref{eq:u_h},\,\ref{eq:sigma_h}) and evaluating the integrals by means of GLL quadrature leads to
\begin{equation}
\label{eq:sigma:matrix form}
\NM M^{\M m} \NM{\MS \sigma}^{\M m}
= \NM M^{\M m} (\NM{\nabla u})^{\M m}
+ \sum_{\gamma=1}^4 \left( \frac{1}{2} + \MS \beta \cdot \M n_{\gamma} \right)
  \NM M^{\Gamma}_{\gamma} (\NM u_{\gamma}^+ - \NM u_{\gamma}^-) \M n_{\gamma}
\, ,
\end{equation}
where
${{\NM{\MS \sigma}^{\M m}} = [ \MS{\sigma}_{ij}^{\M m} ]}$
and
\begin{equation}
\label{eq:grad u}
(\nabla u)_{ij}^{\M m}
= \begin{bmatrix}
     \frac{2}{\Delta x_1^{m_1}} \sum_{p=0}^P D^s_{i,p}\, u_{pj}^{\M m} \\[\medskipamount]
     \frac{2}{\Delta x_2^{m_2}} \sum_{q=0}^P D^s_{j,q}\, u_{iq}^{\M m}
  \end{bmatrix}
  \,
\end{equation}
are the nodal coefficients of $\nabla u_h$ in $\Omega^{\M m}$.
The last term in \eqref{eq:sigma:matrix form} comprises the contributions of the boundary edges.
See Fig.~\ref{fig:element} for illustration and
Tab.~\ref{tab:boundary:vars} for the involved operators and variables.
Multiplying Eq.~\eqref{eq:sigma:matrix form} with the inverse of the element mass matrix yields
\begin{equation}
\label{eq:sigma:matrix form:resolved}
\NM{\MS \sigma}^{\M m}
= (\NM{\nabla u})^{\M m}
+ \sum_{\gamma=1}^4 \left( \frac{1}{2} + \MS \beta \cdot \M n_{\gamma} \right)
  (\NM M^{\M m})^{-1} \NM M^{\Gamma}_{\gamma} (\NM u_{\gamma}^+ - \NM u_{\gamma}^-) \M n_{\gamma}
\, .
\end{equation}
\begin{table}
\renewcommand{\arraystretch}{1.5}
\centering
\caption{Element boundary operators and variables.
\label{tab:boundary:vars}}
\begin{tabular}{c@{\qquad}c@{\qquad}r@{\qquad}c@{\qquad}c@{\qquad}c}
\toprule
$\gamma$ & edge & $\M n_{\gamma}$ & $\NM M^{\Gamma}_{\gamma}$ & $\NM u_{\gamma}^-$ & $\NM u_{\gamma}^+$ \\
\midrule
1 & $\Gamma^{m_1-\sfrac{1}{2},m_2}$ & $-\M e_1$ & $\delta_{i,0}\,M_{2,j,q}^{m_2}$ & $u_{0q}^{m_1,m_2}$ & $u_{Pq}^{m_1-1,m_2}  $ \\
2 & $\Gamma^{m_1,m_2-\sfrac{1}{2}}$ & $-\M e_2$ & $\delta_{j,0}\,M_{1,i,p}^{m_1}$ & $u_{p0}^{m_1,m_2}$ & $u_{pP}^{m_1,  m_2-1}$ \\
3 & $\Gamma^{m_1+\sfrac{1}{2},m_2}$ &  $\M e_1$ & $\delta_{i,P}\,M_{2,j,q}^{m_2}$ & $u_{Pq}^{m_1,m_2}$ & $u_{0q}^{m_1+1,m_2}  $ \\
4 & $\Gamma^{m_1,m_2+\sfrac{1}{2}}$ &  $\M e_2$ & $\delta_{j,P}\,M_{1,i,p}^{m_1}$ & $u_{pP}^{m_1,m_2}$ & $u_{0p}^{m_1,  m_2+1}$ \\
\bottomrule
\end{tabular}
\end{table}
For further simplification it is essential that the boundary operators
$\NM M^{\Gamma}_{\gamma}$ are products of
the injection operator for the normal direction and
the one-dimensional mass matrix \eqref{eq:1D mass matrix} for the tangential direction.
For example, for ${\gamma = 1}$, i.e. $\Gamma^{m_1-\sfrac{1}{2},m_2}$, we find
\begin{align*}
\big[(\NM M^{\M m})^{-1} \NM M^{\Gamma}_{\gamma}\big]_{ij,q}
&= \sum_{k,l=0}^P
     \big(M_{1,i,k}^{m_1}\big)^{-1} \, \big(M_{2,j,l}^{m_2}\big)^{-1} \,
     \delta_{k,0}\,M_{2,l,q}^{m_2}
\\
&= \big(M_{1,i,0}^{m_1}\big)^{-1} \delta_{j,q}
\\
&= \frac{2}{\Delta x_1^{m_1} \rho_0} \,\delta_{i,0}\,\delta_{j,q}
\, .
\end{align*}
Using this result and observing
${\M n_1 = -\M e_1 = -\transpose{[ 1 \, 0 ]}}$,
the contribution of the edge in Eq.~\eqref{eq:sigma:matrix form:resolved} becomes
\begin{equation*}
\left(
\left( \frac{1}{2} + \MS \beta \cdot \M n_{1} \right)
  (\NM M^{\M m})^{-1} \NM M^{\Gamma}_{1} (\NM u_{1}^+ - \NM u_{1}^-) \M n_{1}
\right)_{ij}
= \frac{1 - 2\beta_1}{\Delta x_1^{m_1} \rho_0}
  \delta_{i,0}
  \big(u_{Pj}^{m_1-1,m_2} - u_{0j}^{m_1,m_2} \big)
  \begin{bmatrix*}[r]  -1 \\ 0 \end{bmatrix*}
\, .
\end{equation*}
Substitution of this and the corresponding expressions for ${\gamma = 2, 3, 4}$ in \eqref{eq:sigma:matrix form} yields the flux vector coefficients
\begin{equation}
\label{eq:sigma}
\begin{split}
\MS\sigma_{ij}^{\M m}
&= (\nabla u)_{ij}^{\M m}  \\
&+ \begin{bmatrix}
      \frac{1 - 2\beta_1}{\Delta x_1^{m_1} \rho_0} \delta_{i,0}
      \big( u_{0j}^{m_1,m_2}     - u_{Pj}^{m_1-1,m_2} \big)
    + \frac{1 + 2\beta_1}{\Delta x_1^{m_1} \rho_P} \delta_{i,P}
      \big( u_{0j}^{m_1+1,m_2} - u_{Pj}^{m_1,m_2}     \big)
    \\[\medskipamount]
      \frac{1 - 2\beta_2}{\Delta x_2^{m_2} \rho_0} \delta_{j,0}
      \big( u_{i0}^{m_1,m_2}     - u_{iP}^{m_1,m_2-1} \big)
    + \frac{1 + 2\beta_2}{\Delta x_2^{m_2} \rho_P} \delta_{j,P}
      \big( u_{i0}^{m_1,m_2+1} - u_{iP}^{m_1,m_2}     \big)
  \end{bmatrix}
\, .
\end{split}
\end{equation}

Equations (\ref{eq:grad u},\,\ref{eq:sigma}) immediately lead to explicit expressions for the numerical fluxes $\hat{\MS\sigma}$ defined by \eqref{flux:num:sigma:IP} and \eqref{flux:num:sigma:LDG}, respectively.
As the discrete equations require only the flux normal to the boundary, it is sufficient to consider
the $x_1$-component at edge $\Gamma^{m_1+\sfrac{1}{2},m_2}$
and, correspondingly, the $x_2$-component at $\Gamma^{m_1,m_2+\sfrac{1}{2}}$.
For example, the IP flux \eqref{flux:num:sigma:IP} through $\Gamma^{m_1+\sfrac{1}{2},m_2}$ becomes
\begin{align*}
(\hat{\MS\sigma}_{\textsc{ip}} \cdot \M n)\vert_{\Gamma^{m_1+\sfrac{1}{2},m_2}}
= \M e_1 \cdot \big( \average{\nabla u_h} - \mu_{\textsc{ip}} \jump{u_h} \big)\big\vert_{\Gamma^{m_1+\sfrac{1}{2},m_2}}
\eqqcolon \hat\sigma^{m_1+\sfrac{1}{2},m_2}_{\textsc{ip}, 1}
\; .
\end{align*}
To obtain the nodal coefficients, this expression has to be evaluated at the collocation points.
Expanding the average and jump operators \eqref{eq:trace operators} and substituting \eqref{eq:grad u} yields
\begin{subequations}
\label{flux:num:sigma:components:IP}
\begin{equation}
\begin{aligned}
\hat\sigma^{m_1+\sfrac{1}{2},m_2}_{\textsc{ip},\,1,\,j}
&= \frac{1}{\Delta x_1^{m_1}}   \sum_{k=0}^P D^s_{P,k} u_{kj}^{m_1,m_2}
 + \frac{1}{\Delta x_1^{m_1+1}} \sum_{l=0}^P D^s_{0,l} u_{lj}^{m_1+1,m_2}
\\
&+ \mu_{\textsc{ip}}^{m_1+\sfrac{1}{2},m_2}
     \left(
       u_{0j}^{m_1+1,m_2}
     - u_{Pj}^{m_1,m_2}
     \right)
\end{aligned}
\end{equation}
and, similarly, for the $x_2$-direction
\begin{equation}
\begin{aligned}
\hat\sigma^{m_1,m_2+\sfrac{1}{2}}_{\textsc{ip},\,2,\,i}
&= \frac{1}{\Delta x_2^{m_2}}   \sum_{k=0}^P D^s_{P,k} u_{ik}^{m_1,m_2}
 + \frac{1}{\Delta x_2^{m_2+1}} \sum_{l=0}^P D^s_{0,l} u_{il}^{m_1,m_2+1}
\\
&+ \mu_{\textsc{ip}}^{m_1,m_2+\sfrac{1}{2}}
     \left(
       u_{i0}^{m_1,m_2+1}
     - u_{iP}^{m_1,m_2}
     \right).
\end{aligned}
\end{equation}
\end{subequations}
The corresponding LDG flux is
\begin{align*}
(\hat{\MS\sigma}_{\textsc{ldg}} \cdot \M n)\vert_{\Gamma^{m_1+\sfrac{1}{2},m_2}}
= \M e_1 \cdot
  \big( \average{\MS\sigma_h}
      + \MS\beta\jump{\MS\sigma_h}
      - \mu_{\textsc{ldg}} \jump{u_h}
  \big)\big\vert_{\Gamma^{m_1+\sfrac{1}{2},m_2}}
\eqqcolon \hat\sigma^{m_1+\sfrac{1}{2},m_2}_{\textsc{ldg}, 1}
\; .
\end{align*}
With \eqref{eq:trace operators} and \eqref{eq:sigma} this leads to the nodal coefficients
\begin{subequations}
\label{flux:num:sigma:components:LDG}
\begin{equation}
\begin{aligned}
\hat\sigma^{m_1+\sfrac{1}{2},m_2}_{\textsc{ldg},\,1,\,j}
&= \frac{1 + 2\beta_1}{\Delta x_1^{m_1}}   \sum_{k=0}^P D^s_{P,k} u_{kj}^{m_1,m_2}
 + \frac{1 - 2\beta_1}{\Delta x_1^{m_1+1}} \sum_{l=0}^P D^s_{0,l} u_{lj}^{m_1+1,m_2}
\\
&+ \left[
     \frac{(1 + 2\beta_1)^2}{2\Delta x_1^{m_1} \rho_P}
   + \frac{(1 - 2\beta_1)^2}{2\Delta x_1^{m_1+1} \rho_0}
   + \mu_{\textsc{ldg}}^{m_1+\sfrac{1}{2},m_2}
   \right]
   \left(
     u_{0j}^{m_1+1,m_2}
   - u_{Pj}^{m_1,m_2}
   \right).
\end{aligned}
\end{equation}
For direction 2 we obtain
\begin{equation}
\begin{aligned}
\hat\sigma^{m_1,m_2+\sfrac{1}{2}}_{\textsc{ldg},\,2,\,i}
&= \frac{1 + 2\beta_2}{\Delta x_2^{m_2}}   \sum_{k=0}^P D^s_{P,k} u_{ik}^{m_1,m_2}
 + \frac{1 - 2\beta_2}{\Delta x_2^{m_2+1}} \sum_{l=0}^P D^s_{0,l} u_{il}^{m_1,m_2+1}
\\
&+ \left[
     \frac{(1 + 2\beta_2)^2}{2\Delta x_2^{m_2} \rho_P}
   + \frac{(1 - 2\beta_2)^2}{2\Delta x_2^{m_2+1} \rho_0}
   + \mu_{\textsc{ldg}}^{m_1,\,m_2+\sfrac{1}{2}}
   \right]
   \left(
     u_{i0}^{m_1,m_2+1}
   - u_{iP}^{m_1,m_2}
   \right).
\end{aligned}
\end{equation}
\end{subequations}
A closer inspection of Equations (\ref{eq:sigma}--\ref{flux:num:sigma:components:LDG}) shows
that IP and LDG coincide if \mbox{$\MS\beta = 0$} and
\begin{align*}
\mu_{\textsc{ip}}^{m_1+\sfrac{1}{2},m_2}
&= \left( \frac{1}{2\Delta x_1^{m_1} \rho_P} + \frac{1}{2\Delta x_1^{m_1+1} \rho_0} \right)
 + \mu_{\textsc{ldg}}^{m_1+\sfrac{1}{2},m_2}
 = \mu_{0}^{m_1+\sfrac{1}{2},m_2}
 + \mu_{\textsc{ldg}}^{m_1+\sfrac{1}{2},m_2}
,
\\
\mu_{\textsc{ip}}^{m_1,\,m_2+\sfrac{1}{2}}
&= \left( \frac{1}{2\Delta x_2^{m_2} \rho_P} + \frac{1}{2\Delta x_2^{m_2+1} \rho_0} \right)
 + \mu_{\textsc{ldg}}^{m_1,\,m_2+\sfrac{1}{2}}
 = \mu_{0}^{m_1,\,m_2+\sfrac{1}{2}}
 + \mu_{\textsc{ldg}}^{m_1,\,m_2+\sfrac{1}{2}}
.
\end{align*}
This observation allows to relate the interior penalty method to the LDG stability condition
\mbox{$\mu_{\textsc{ldg}} > 0$} and motivates the generic penalty coefficient
\begin{gather*}
\mu = \mu_0 \, (1 + \mu_{\star}), \quad
\end{gather*}
where \mbox{$\mu_{\star} > 0$} is a dimensionless parameter and $\mu_0$ the LDG stability threshold, e.g., at $\Gamma^{m_1+\sfrac{1}{2},m_2}$
\begin{equation*}
\mu_0^{m_1+\sfrac{1}{2},m_2}
= \frac{1}{2\Delta x_1^{m_1} \rho_P} + \frac{1}{2\Delta x_1^{m_1+1} \rho_0}
= \left\{\frac{1}{\Delta x_1}\right\} \frac{P(P+1)}{2}
\, .
\end{equation*}
The original coefficients are related to the generic one by
\mbox{$\mu_{\textsc{ip}} = \mu$} and
\mbox{$\mu_{\textsc{ldg}} = \mu - \mu_0$}.
As the penalty coefficients depend only on the mesh spacing normal to the edge, we simplify the notation by dropping the index referring to the tangential direction, i.e. $\mu^{m_1+\sfrac{1}{2},m_2}$ becomes $\mu^{m_1+\sfrac{1}{2}}$ etc.
Introducing these definitions in \eqref{flux:num:sigma:components:IP} and \eqref{flux:num:sigma:components:LDG} yields the unified numerical fluxes
\begin{subequations}
\label{eq:flux:num:sigma}
\begin{align}
\begin{split}
\label{eq:flux:num:sigma:x}
\hat\sigma^{m_1+\sfrac{1}{2},m_2}_{1,j}
&= \frac{1 + 2\beta_1}{\Delta x_1^{m_1}}   \sum_{k=0}^P D^s_{P,k} u_{kj}^{m_1,m_2}
 + \frac{1 - 2\beta_1}{\Delta x_1^{m_1+1}} \sum_{l=0}^P D^s_{0,l} u_{lj}^{m_1+1,m_2}
\\
&+ \left[
     2\frac{\beta_1^2 + \beta_1}{\Delta x_1^{m_1} \rho_P}
   + 2\frac{\beta_1^2 - \beta_1}{\Delta x_1^{m_1+1} \rho_0}
   + \mu^{m_1+\sfrac{1}{2}}
   \right]
   \left(
     u_{0j}^{m_1+1,m_2}
   - u_{Pj}^{m_1,m_2}
   \right),
\end{split}
\\[\medskipamount]
\begin{split}
\label{eq:flux:num:sigma:y}
\hat\sigma^{m_1,m_2+\sfrac{1}{2}}_{2,\,i}
&= \frac{1 + 2\beta_2}{\Delta x_2^{m_2}}   \sum_{k=0}^P D^s_{P,k} u_{ik}^{m_1,m_2}
 + \frac{1 - 2\beta_2}{\Delta x_2^{m_2+1}} \sum_{l=0}^P D^s_{0,l} u_{il}^{m_1,m_2+1}
\\
&+ \left[
     2\frac{\beta_2^2 + \beta_2}{\Delta x_2^{m_2} \rho_P}
   + 2\frac{\beta_2^2 - \beta_2}{\Delta x_2^{m_2+1} \rho_0}
   + \mu^{m_2+\sfrac{1}{2}}
   \right]
   \left(
     u_{i0}^{m_1,m_2+1}
   - u_{iP}^{m_1,m_2}
     \right).
\end{split}
\end{align}
\end{subequations}

With these prerequisites we are ready to tackle \eqref{eq:dg:elemental:2}.
To obtain the discrete equations we set
${v|_{\Omega^{\M m}} = \varphi_i(\xi_1) \varphi_j(\xi_2)}$
and evaluate each integral by means of GLL quadrature.
The first term of \eqref{eq:dg:elemental:2} becomes
\begin{equation*}
\int_{\Omega^{\M m}} \!\! \nabla v \cdot \MS{\sigma}_h \,\D\Omega
= \frac{\Delta x_1^{m1} \Delta x_2^{m_2}}{4}
  \sum_{p,q=0}^P \rho_p \rho_q \MS \sigma_{pq}^{\M m} \cdot
                 \begin{bmatrix*}
                   \frac{2}{\Delta x_1^{m_1}}  \varphi_i'(\eta_p)  \varphi_j(\eta_q) \\
                   \frac{2}{\Delta x_2^{m_2}}  \varphi_i(\eta_p)   \varphi_j'(\eta_q)
                 \end{bmatrix*}
\, ,
\end{equation*}
where $\eta_p$ and $\eta_q$ denote the GLL points for direction 1 and 2, respectively.
Using the already introduced 1D mass and differentiation matrices, and exploiting the orthogonality of the former, this can be rewritten to
\begin{equation*}
\int_{\Omega^{\M m}} \!\! \nabla v \cdot \MS{\sigma}_h \,\D\Omega
= M_{2,jj}^{m_2} \sum_{p=0}^P \rho_p D^s_{p,i} \sigma_{1,pj}^{\M m}
+ M_{1,ii}^{m_1} \sum_{q=0}^P \rho_q D^s_{q,j} \sigma_{2,iq}^{\M m}
\, .
\end{equation*}
Substitution of
$\sigma^{\M m}_{1,pj}$ and $\sigma^{\M m}_{2,iq}$ with \eqref{eq:sigma} and, therein,
$(\nabla u)_{ij}^{\M m}$ with \eqref{eq:grad u} gives
\begin{equation*}
\begin{alignedat}{2}
\int_{\Omega^{\M m}} \!\! \nabla v \cdot \MS{\sigma}_h \,\D\Omega
&= M_{2,j,j}^{m_2}
    \sum_{p=0}^P
      \rho_p D^s_{p,i} \Big[ \cramped{%
                           \frac{2}{\Delta x_1^{m_1}} \sum_{k=0}^P D^s_{p,k}\, u_{kj}^{\M m}
                           }
                            && + \cramped{%
                                 \frac{1 - 2\beta_1}{\Delta x_1^{m_1} \rho_0} \delta_{p,0}
                                 \big( u_{0j}^{m_1,m_2} - u_{Pj}^{m_1-1,m_2} \big)
                                 }
\\
&                           && + \cramped{%
                                 \frac{1\! + 2\beta_1}{\Delta x_1^{m_1} \rho_P} \delta_{p,P}
                                 \big( u_{0j}^{m_1+1,m_2} - u_{Pj}^{m_1,m_2} \big)
                                 }
                     \Big]
\\
&+ M_{1,i,i}^{m_1}
    \sum_{q=0}^P
      \rho_q D^s_{q,j} \Big[ \cramped{%
                           \frac{2}{\Delta x_2^{m_2}} \sum_{l=0}^P D^s_{p,l}\, u_{il}^{\M m}
                           }
                            && + \cramped{%
                                 \frac{1 - 2\beta_2}{\Delta x_2^{m_2} \rho_0} \delta_{q,0}
                                 \big( u_{i0}^{m_1,m_2} - u_{iP}^{m_1,m_2-1} \big)
                                 }
\\
&                           && + \cramped{%
                                 \frac{1\! + 2\beta_2}{\Delta x_2^{m_2} \rho_P} \delta_{q,P}
                                 \big( u_{i0}^{m_1,m_2+1} - u_{iP}^{m_1,m_2} \big)
                                 }
                     \Big]
\, .
\end{alignedat}
\end{equation*}
Introduction of the one-dimensional standard stiffness matrix
\begin{equation*}
L_{i,k}^{s}
= \int_{-1}^{1} \varphi_i'(\xi)  \varphi_k'(\xi) \D \xi
= \sum_{j=0}^{P} \rho_j D^s_{j,i} D^s_{j,k}
\end{equation*}
and exploitation of the Kronecker deltas yields the final form
\begin{equation*}
\begin{alignedat}{2}
\int_{\Omega^{\M m}} \!\! \nabla v \cdot \MS{\sigma}_h \,\D\Omega
&= M_{2,j,j}^{m_2}
    \Big[ \cramped{%
          \frac{1}{\Delta x_1^{m_1}} \sum_{k=0}^P L_{i,k}^{s} u_{pj}^{\M m}
          }
          && + \cramped{%
               \frac{1 - 2\beta_1}{\Delta x_1^{m_1}} D^s_{0,i}
               \big( u_{0j}^{m_1,m_2} - u_{Pj}^{m_1-1,m_2} \big)
               }
\\
&         && + \cramped{%
               \frac{1 + 2\beta_1}{\Delta x_1^{m_1}}  D^s_{P,i}
               \big( u_{0j}^{m_1+1,m_2} - u_{Pj}^{m_1,m_2} \big)
               }
     \Big]
\\
&+ M_{1,i,i}^{m_1}
    \Big[ \cramped{%
          \frac{1}{\Delta x_2^{m_2}} \sum_{l=0}^P L_{j,l}^{s} u_{il}^{\M m}
          }
          && + \cramped{%
               \frac{1 - 2\beta_2}{\Delta x_2^{m_2}} D^s_{0,j}
               \big( u_{i0}^{m_1,m_2} - u_{iP}^{m_1,m_2-1} \big)
               }
\\
&         && + \cramped{%
               \frac{1 + 2\beta_2}{\Delta x_2^{m_2}}  D^s_{P,j}
               \big( u_{i0}^{m_1,m_2+1} - u_{iP}^{m_1,m_2} \big)
               }
     \Big]
\, .
\end{alignedat}
\end{equation*}
The second term in Eq.~\eqref{eq:dg:elemental:2} becomes
\begin{equation*}
\int_{\Omega^{\M m}} \!\! v f \,\D\Omega
= \frac{\Delta x_1^{m_1} \Delta x_2^{m_2}}{4} \, \rho_i \rho_j \,
   f\big(x_1^{m_1}(\eta_i), x_2^{m_2}(\eta_j)\big)
 \eqqcolon g_{ij}^{m_1,m_2}
\, ,
\end{equation*}
where
\mbox{$x_d^{m_d}(\cdot)$}
is the inverse element mapping.
Finally, the last term represents an integral over the element boundary, which can be decomposed into contributions of the four edges:
\begin{equation*}
\begin{split}
\int_{\partial\Omega^{\M m}} \!\! v \hat{\MS\sigma} \cdot \M n \,\D\Gamma
= & - \int_{\Gamma^{m_1-\sfrac{1}{2},m_2}} \! v \hat\sigma_1 \,\D\Gamma
    - \int_{\Gamma^{m_1,m_2-\sfrac{1}{2}}} \! v \hat\sigma_2 \,\D\Gamma \\
  & + \int_{\Gamma^{m_1+\sfrac{1}{2},m_2}} \! v \hat\sigma_1 \,\D\Gamma
    + \int_{\Gamma^{m_1,m_2+\sfrac{1}{2}}} \! v \hat\sigma_2 \,\D\Gamma
\, .
\end{split}
\end{equation*}
Applying GLL quadrature on the edges yields
\begin{equation*}
\begin{split}
\int_{\partial\Omega^{\M m}} \!\! v \hat{\MS\sigma} \cdot \M n \,\D\Gamma
= & - \delta_{i,0} M_{2,j,j}^{m_2} \hat\sigma_{1,j}^{m_1-\sfrac{1}{2},m_2}
    - \delta_{j,0} M_{1,i,i}^{m_1} \hat\sigma_{2,i}^{m_1,m_2-\sfrac{1}{2}} \\
  & + \delta_{i,P} M_{2,j,j}^{m_2} \hat\sigma_{1,j}^{m_1+\sfrac{1}{2},m_2}
    + \delta_{j,P} M_{1,i,i}^{m_1} \hat\sigma_{2,i}^{m_1,m_2+\sfrac{1}{2}}
\, .
\end{split}
\end{equation*}
The terms on the right side can be expanded using the expressions for the numerical fluxes given in
\eqref{eq:flux:num:sigma}.
Doing this and substituting the above results in \eqref{eq:dg:elemental:2} leads to the discrete equations
\begin{alignat}{2}
& M_{jj}^{m_2}
   && \bigg[ \frac{2}{\Delta x_1^{m_1}} \sum_{k=0}^P L^s_{i,k} u_{kj}^{m_1,m_2}
\nonumber
\\
&  && + \frac{1 - 2\beta_1}{\Delta x_1^{m_1}} D^s_{0,i} \big( u_{0j}^{m_1,m_2} - u_{Pj}^{m_1-1,m_2} \big)
\nonumber
\\
&  && + \frac{1 + 2\beta_1}{\Delta x_1^{m_1}} D^s_{P,i} \big( u_{0j}^{m_1+1,m_2} - u_{Pj}^{m_1,m_2} \big)
\nonumber
\\
&  && \begin{aligned}
        +\; \delta_{i,0}
            \bigg(
              \, &
              \Big(
                    \frac{1 + 2\beta_1}{\Delta x_1^{m_1-1}} \sum_{k=0}^P D^s_{P,k} u_{kj}^{m_1-1,m_2}
                  + \frac{1 - 2\beta_1}{\Delta x_1^{m_1}}   \sum_{l=0}^P D^s_{0,l} u_{lj}^{m_1,m_2}
              \Big)
\\
              +\, &
              \Big(
                    2\frac{\beta_1^2 + \beta_1}{\Delta x_1^{m_1-1} \rho_P}
                  + 2\frac{\beta_1^2 - \beta_1}{\Delta x_1^{m_1}   \rho_0}
                  + \mu^{m_1-\sfrac{1}{2}}
              \Big)
              \Big(
                    u_{0j}^{m_1,m_2}
                  - u_{Pj}^{m_1-1,m_2}
              \Big)
            \bigg)
        \end{aligned}
\nonumber
\\
&  && \begin{aligned}
        -\; \delta_{i,P}
            \bigg(
              \, &
              \Big(
                      \frac{1 + 2\beta_1}{\Delta x_1^{m_1}}   \sum_{k=0}^P D^s_{P,k} u_{kj}^{m_1,m_2}
                    + \frac{1 - 2\beta_1}{\Delta x_1^{m_1+1}} \sum_{l=0}^P D^s_{0,l} u_{lj}^{m_1+1,m_2}
              \Big)
\\
              +\, &
              \Big(
                      2\frac{\beta_1^2 + \beta_1}{\Delta x_1^{m_1} \rho_P}
                    + 2\frac{\beta_1^2 - \beta_1}{\Delta x_1^{m_1+1} \rho_0}
                    + \mu^{m_1+\sfrac{1}{2}}
              \Big)
              \Big(
                      u_{0j}^{m_1+1,m_2}
                    - u_{Pj}^{m_1,m_2}
              \Big)
            \bigg)
          \bigg]
        \end{aligned}
\nonumber
\\[\smallskipamount]
+~& M_{ii}^{m_1}
   && \bigg[ \frac{2}{\Delta x_2^{m_2}} \sum_{l=0}^P L^s_{j,l} u_{il}^{m_1,m_2}
\nonumber
\\
&  && + \frac{1 - 2\beta_2}{\Delta x_2^{m_2}} D^s_{0,j} \big( u_{i0}^{m_1,m_2} - u_{iP}^{m_1,m_2-1} \big)
\nonumber
\\
&  && + \frac{1 + 2\beta_2}{\Delta x_2^{m_2}} D^s_{P,j} \big( u_{i0}^{m_1,m_2+1} - u_{iP}^{m_1,m_2} \big)
\nonumber
\\
&  && \begin{aligned}
        +\; \delta_{j,0}
            \bigg(
              \, &
              \Big(
                      \frac{1 + 2\beta_2}{\Delta x_2^{m_2-1}} \sum_{k=0}^P D^s_{P,k} u_{ik}^{m_1,m_2-1}
                    + \frac{1 - 2\beta_2}{\Delta x_2^{m_2}}   \sum_{l=0}^P D^s_{0,l} u_{il}^{m_1,m_2}
              \Big)
\\
              +\, &
              \Big(
                      2\frac{\beta_2^2 + \beta_2}{\Delta x_2^{m_2-1} \rho_P}
                    + 2\frac{\beta_2^2 - \beta_2}{\Delta x_2^{m_2}   \rho_0}
                    + \mu^{m_2-\sfrac{1}{2}}
              \Big)
              \Big(
                      u_{i0}^{m_1,m_2}
                    - u_{iP}^{m_1,m_2-1}
              \Big)
            \bigg)
        \end{aligned}
\nonumber
\\
&  && \begin{aligned}
        -\; \delta_{j,P}
            \bigg(
              \, &
              \Big(
                      \frac{1 + 2\beta_2}{\Delta x_2^{m_2}}   \sum_{k=0}^P D^s_{P,k} u_{ik}^{m_1,m_2}
                    + \frac{1 - 2\beta_2}{\Delta x_2^{m_2+1}} \sum_{l=0}^P D^s_{0,l} u_{il}^{m_1,m_2+1}
              \Big)
\\
              +\, &
              \Big(
                      2\frac{\beta_2^2 + \beta_2}{\Delta x_2^{m_2} \rho_P}
                    + 2\frac{\beta_2^2 - \beta_2}{\Delta x_2^{m_2+1} \rho_0}
                    + \mu^{m_2+\sfrac{1}{2}}
              \Big)
              \Big(
                      u_{i0}^{m_1,m_2+1}
                    - u_{iP}^{m_1,m_2}
              \Big)
            \bigg)
          \bigg]
        \end{aligned}
\nonumber
\\[\smallskipamount]
=~& g_{ij}^{m_1,m_2} \hspace*{-4em}
\label{eq:primal form}
\end{alignat}
for
\mbox{$0 \le i, j \le P$} and
\mbox{$1 \le m_d \le \nel[,d]$}.

To emphasize the tensor-product structure of \eqref{eq:primal form} we introduce the global solution coefficients $u_{IJ} = u_{ij}^{m_1,m_2}$
with periodic indices
\begin{subequations}
\label{eq:IJ}
\begin{align}
I &= \ell_1(i,m_1) \coloneqq i + (P+1) \tilde (m_1 - 1) \\
J &= \ell_2(j,m_2) \coloneqq j + (P+1) \tilde (m_2 - 1) \, ,
\end{align}
\end{subequations}
where
${\tilde m_d = m_d + k \nel[,d]}$
is the 1D element index mapped into the range
\mbox{$1 \le \tilde m_d \le \nel[,d]$}
by properly choosing
${k \in \mathbb{I}}$.
Further, we define the global 1D mass matrices
\begin{equation}
\label{eq:M:generic}
M_{d,I,K} = M_{d,\,\ell_d(i,m),\,\ell_d(k,m+r)} = M^m_{d,i,k} \, \delta_{0,r}
\end{equation}
with ${d=1,2}$ for directions 1 and 2.
%
%
%
The bracketed expressions in \eqref{eq:primal form} can be represented by means of global 1D stiffness matrices that are applied to the solution coefficients.
Careful examination of the expressions shows that these matrices possess the form
\begin{equation}
\label{eq:L:generic}
L_{d,I,K} =
L_{\ell_d(i,m),\,\ell_d(k,m+r)} =
\left\{
  \begin{array}{l@{\qquad}c@{~\,}c@{~}r}
    L_{d,i,k}^-   &  r  & = & -1  \\[\smallskipamount]
    L_{d,i,k}^0   &  r  & = &  0  \\[\smallskipamount]
    L_{d,i,k}^+   &  r  & = &  1  \\[\smallskipamount]
    0           & |r| & > &  1
  \end{array}
\right.
\end{equation}
with
\begin{alignat*}{4}
L_{d,i,k}^0
&= &\phantom{+}& \frac{2}{\Delta x_d^{m}} L^s_{i,k}
\\
& &~+~& \cramped{\frac{1 - 2\beta_d}{\Delta x_d^{m}} D^s_{0,i} \delta_{k,0}}
  &~+~& \cramped{\frac{1 - 2\beta_d}{\Delta x_d^{m}} \delta_{i,0} D^s_{0,k}}
  &~+~& \cramped{
          \left[
               2\frac{\beta_d^2 + \beta_d}{\Delta x_d^{m-1} \rho_P}
             + 2\frac{\beta_d^2 - \beta_d}{\Delta x_d^{m}   \rho_0}
             + \mu^{m-\sfrac{1}{2}}
          \right]
          \delta_{i,0} \delta_{k,0}
        }
\\
& &~-~& \cramped{\frac{1 + 2\beta_d}{\Delta x_d^{m}} D^s_{\!P,i} \delta_{k,P}}
  &~-~& \cramped{\frac{1 + 2\beta_d}{\Delta x_d^{m}} \delta_{i,P} D^s_{\!P,k}}
  &~+~& \cramped{
          \left[
               2\frac{\beta_d^2 + \beta_d}{\Delta x_d^{m}   \rho_P}
             + 2\frac{\beta_d^2 - \beta_d}{\Delta x_d^{m+1} \rho_0}
             + \mu^{m+\sfrac{1}{2}}
          \right]
          \delta_{i,P} \delta_{k,P}
        }
\\[\medskipamount]
L_{d,i,k}^-
&= &~-~& \cramped{\frac{1 - 2\beta_d}{\Delta x_d^{m-1}} D^s_{0,i} \delta_{k,P}}
   &~+~& \cramped{\frac{1 + 2\beta_d}{\Delta x_d^{m-1}} \delta_{i,0} D^s_{\!P,k}}
   &~-~& \cramped{
           \left[
                2\frac{\beta_d^2 + \beta_d}{\Delta x_d^{m-1} \rho_P}
              + 2\frac{\beta_d^2 - \beta_d}{\Delta x_d^{m}   \rho_0}
              + \mu^{m-\sfrac{1}{2}}
           \right]
           \delta_{i,0} \delta_{k,P}
         }
\\[\medskipamount]
L_{d,i,k}^+
&= &\,\phantom{+}\,&
         \cramped{\frac{1 + 2\beta_d}{\Delta x_d^{m+1}} D^s_{\!P,i} \delta_{k,0}}
   &~-~& \cramped{\frac{1 - 2\beta_d}{\Delta x_d^{m+1}} \delta_{i,P} D^s_{0,k}}
   &~-~& \cramped{
           \left[
                2\frac{\beta_d^2 + \beta_d}{\Delta x_d^{m}   \rho_P}
              + 2\frac{\beta_d^2 - \beta_d}{\Delta x_d^{m+1} \rho_0}
              + \mu^{m+\sfrac{1}{2}}
           \right]
           \delta_{i,P} \delta_{k,0}
         }
\,,
\end{alignat*}
where it should be noted that $\mu^{m\pm\sfrac{1}{2}}$ depends on
$\Delta x_d^{m}$ and $\Delta x_d^{m\pm1}$.
Adopting the global 1D operators allows to rewrite
\eqref{eq:primal form} as
\begin{equation}
\label{eq:discrete problem:components}
\sum_{L=0}^{N_2} \sum_{K=0}^{N_1} \,
 \underbrace{%
   \left( M_{2,J,L} L_{1,I,K}
        + L_{2,J,L} M_{1,I,K}
   \right)
   }_{\eqqcolon A_{IJ,KL}}
   u_{KL}
= g_{IJ}
\end{equation}
for
\mbox{$0 \le I \le N_1 = \np\nel[,1]$} and
\mbox{$0 \le J \le N_2 = \np\nel[,2]$},
where ${\np = P+1}$.
Alternatively, the system can be written in matrix form as follows:
\begin{equation}
\label{eq:discrete problem}
\NM A\, \NM u
= (\NM M_2 \otimes \NM L_1 + \NM L_2 \otimes \NM M_1)\, \NM u
= \NM{g}
\,.
\end{equation}
%


\section{Solution methods}
\label{sec:methods}

The linear system \eqref{eq:discrete problem} is symmetric positive semi-definite.
Moreover, its structure closely resembles the discrete equations generated with the continuous spectral element method \cite{LF05,Stiller2016jsc}.
This coincidence inspired us to adopt the multigrid techniques developed in \cite{Stiller2016jsc} for the present discontinuous formulation.
In particular, we examine polynomial multigrid (MG) and multigrid-preconditioned conjugate gradients (MGCG).
Both approaches employ overlapping Schwarz methods for smoothing.
We first present the Schwarz methods and then sketch MG and MGCG.


\subsection{Schwarz methods}
\label{sec:Schwarz}

\begin{figure}
\centering
\hspace*{-2em}
\subfloat[Element-centered subdomain.]
  {\hspace*{1em}
   \includegraphics[scale=0.22]{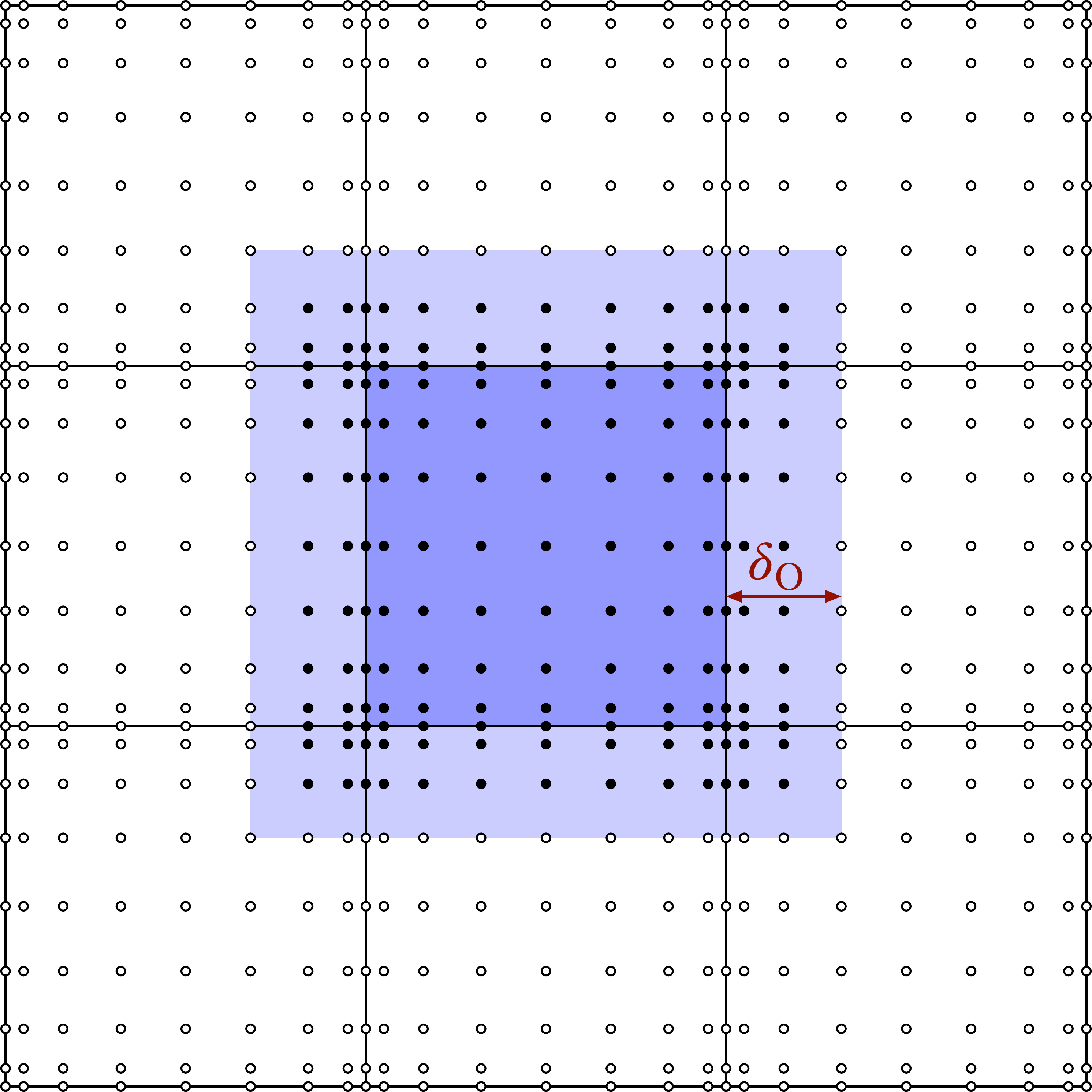}
   \hspace*{1em}
   \label{fig:schwarz:domain:ec}} 
\hspace*{-2em}
\subfloat[Face-centered subdomain for an $x_1$-face.]
  {\hspace*{3.25em}
   \includegraphics[scale=0.22]{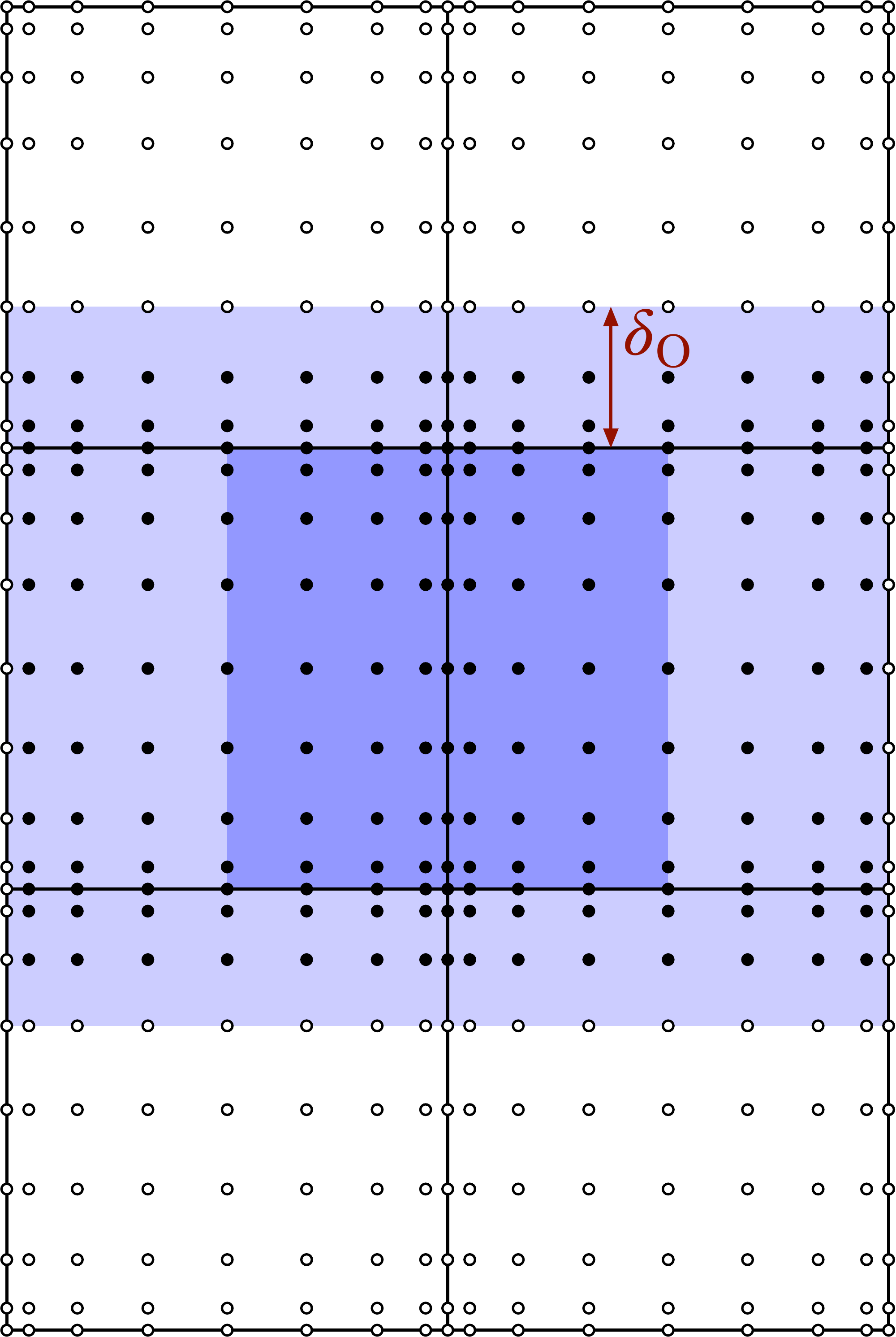}
   \hspace*{2.75em}
   \label{fig:schwarz:domain:fc}} 
\hspace*{-3.75em}
\caption{Subdomains used with the Schwarz method.
Each subdomain consists of a core region (dark shaded) and an overlap zone of width $\deltaO$ (light shaded).
The circles are the GLL nodes for polynomial order $p=8$. Filled circles indicate the nodes that are solved for and updated.
\label{fig:schwarz:domains}}
\end{figure}

Schwarz methods are iterative domain decomposition techniques which improve the approximate solution by parallel or sequential subdomain solves, leading to additive or multiplicative methods, respectively.
Here we consider element-centered and face-centered subdomains as illustrated in Fig.~\ref{fig:schwarz:domains}.
The element-centered subdomain was already used in \cite{Stiller2016jsc}.
It covers the element region extended by a strip including $\no$ layers of additional nodes from the neighbor elements, excluding the nodes located on the subdomain boundary (Fig.~\ref{fig:schwarz:domain:ec}).
The overlap is defined as
\begin{equation*}
\deltaO = \frac{1}{2}\Delta\xi_{\textsc o}\Delta x_n
\, ,
\end{equation*}
where
$\Delta\xi_{\textsc o}$ is the overlap in standard coordinates and
$\Delta x_n$ the extension of the abutting element normal to the shared edge.
For the element-centered subdomain, we set
${\Delta\xi_{\textsc o} = \eta_{\no} + 1}$.
With this definition, the overlap width corresponds to the distance between the edge and the nearest node layer that is not included.
In the case ${\no=0}$ no nodes are adopted and the overlap width equals zero, since ${\eta_{\no} = -1}$.
Choosing ${\no=1}$ includes the boundary nodes of the adjoining elements that are located on the common edge or vertex.
Note, however, that these nodes are counted separately, though they coincide geometrically with nodes of the element representing the core of the subdomain.
Except for the non-overlapping case, the use of element-centered subdomains implies a diagonal coupling, which is natural to continuous elements, but seems artificial to DG.
This motivated the use of face-centered subdomains, including only nodes from the two elements sharing one face (i.e.\ one edge in 2D).
However, in course of our studies it proved necessary to allow for a lateral overlap analogous to the element-centered approach, which finally led to the face-centered domains sketched in Fig.~\ref{fig:schwarz:domain:fc}.
The core region of the face-centered subdomain is defined as the union of the two adjoining half elements, and the opposite halves form a part of the overlap zone. This ``normal'' overlap is fixed and has always an extension of half the element width.
In contrast, the tangential or ``lateral'' overlap is adjustable and defined as in the element-centered case.
According to the face orientation, the subdomains can be divided into two groups, where the normal is aligned either with the $x_1$- or the $x_2$-direction.
In terms of core regions, each group represents a complete partition of the computational domain and, hence, will be treated separately.

In the following ${\{\Omega_s\}}_{s=1}^{\nsd}$ denotes the set of subdomains constituting a single partition, i.e.
the set of element-centered subdomains, or
the set of face-centered subdomains with normals oriented in direction 1, or
the set of face-centered subdomains with normals oriented in direction 2.
Within each set the subdomains are linearly numbered according to a lexicographical ordering based on their row and column indices.
Note that for periodic Cartesian grids ${\nsd=\nel}$, i.e., the number of subdomains equals the number of elements in all three cases.

The main idea of the Schwarz method is
to solve a subproblem for every subdomain $\Omega_s$ and to construct a correction to a given approximate solution $\NM{\tilde u}$ by combining the resulting local corrections.
For establishing the subproblems we introduce the (exact) correction
${\Delta \NM u = \NM u -  \NM{\tilde u}}$
and convert Eq.~\eqref{eq:discrete problem} into the equivalent residual form
\begin{equation*}
\NM A \Delta \NM u = \NM{g} - \NM A \NM{\tilde u} = \NM{\tilde r} \, .
\end{equation*}
For each subdomain $\Omega_{s}$
we define the restriction operator $\NM R_s$ such that
${\NM u_s = \NM R_s \NM u}$ gives the associated coefficients.
Conversely, the transposed restriction operator, $\transpose{\NM R_s}$ is used to globalize the local coefficients by adding zeros for exterior nodes.
With these prerequisites the correction contributed by $\Omega_{s}$ is defined as the solution of the subproblem
\begin{equation}
\label{eq:subproblem}
\NM A_{ss} \Delta \NM u_s = \NM r_s \, ,
\end{equation}
where
\mbox{$\NM A_{ss} = \NM R_s \NM A \, \transpose{\NM R}_s$}
represents the restricted system matrix and
\mbox{$\NM r_s = \NM R_s \NM{\tilde r}$}
the restricted residual.
Due to the rectangular shape of the subdomain, the restriction operator possesses the tensor-product factorization
${\NM R_s = \NM R_{s,1} \otimes \NM R_{s,2}}$
and $\NM A_{ss}$ inherits the structure of the full system matrix $\NM A$, i.e.
\begin{equation*}
\NM A_{ss} = \NM M_{s,2} \otimes \NM L_{s,1} + \NM L_{s,2} \otimes \NM M_{s,1}
\, .
\end{equation*}
Note that the one-dimensional mass matrices
${\NM M_{s,d} = \NM R_{s,d} \NM M_d} \transpose{\NM R_{s,d}}$
are positive diagonal,
while the stiffness matrices
${\NM L_{s,d} = \NM R_{s,d} \NM L_d \transpose{\NM R_{s,d}}}$
are symmetric and --- because of the implied Dirichlet conditions --- positive definite.
These properties allow for application of the fast diagonalization technique (FDM) developed by \citet{Lyn64} and adopted for SEM in \cite{CD95}, which yields the inverse of $\NM A_{ss}$ in the form
\begin{equation*}
\NM A_{ss}^{-1}
= (\NM S_2 \otimes \NM S_1)
  (\NM I \otimes \NM \varLambda_1 + \NM \varLambda_2 \otimes \NM I)^{-1}
  (\transpose{\NM S_2} \otimes \transpose{\NM S_1}),
\end{equation*}
where $\NM I$ is the unity matrix,
$\NM S_d$ a matrix composed of the eigenvectors to the generalized eigenproblem for $\NM L_{s,d}$ and $\NM M_{s,d}$, and
$\NM \varLambda_d$ the diagonal matrix of eigenvalues.
The central term on the right side is the diagonal matrix composed of the reciprocal 2D eigenvalues.
Apart from this factor, the evaluation of
${\NM A_{ss}^{-1} \NM r_s}$ requires the subsequent application of four 1D operators, namely
$\NM{S}_1$, $\NM{S}_2$ and their transposes, to a 2D operand.
For an element-centered domain,
$\NM{S}_1$ and $\NM{S}_2$ are square matrices of dimension ${\np + 2\no}$, where
${\np=P+1}$ is the number of collocation points in one direction, and
$\NM r_s$ is of the size $({\np + 2\no)^2}$.
Hence, ${\Delta\NM u_s = \NM A_{ss}^{-1} \NM r_s}$
can be computed with ${\Theta(4(\np+\no)^3)}$ operations.
In the face-centered case, $\NM{S}_1$ and $\NM{S}_2$ are square matrices of dimensions
${2\np-2}$ and ${\np + 2\no}$, or vice versa, depending on the face orientation, and
$\NM r_s$ is of the size ${(2\np-2)(\np + 2\no)}$.
Assuming ${2\np-2 \approx 2\np}$, the cost for evaluating $\NM u_s$ amounts to
${4(3\np+2\no)(\np+2\no)\np}$.
Finally, defining
  ${\cd =  4 (1 + 2\co)^3}$                      for element-centered and
  ${\cd = 12 (1 + 2\co) (1 + \sfrac{2\co}{3})}$  for face-centered subdomains,
where
  ${\co = \sfrac{\no}{\np}}$,
both estimates can be cast in the form ${\Theta(\cd\np^3)}$.

Several options exist for combining the local solutions.
Following \cite{Stiller2016jsc} we consider the multiplicative Schwarz method and a weighted version of the additive Schwarz method.
The multiplicative Schwarz method solves the subproblems \eqref{eq:subproblem} consecutively, while continually updating the residual.
In the element-centered case, one iteration corresponds to one sweep over all subdomains, whereas two sweeps are performed in the face-centered case: one for over the domains oriented in direction 1 and another one for direction 2.
Though $\NM A$ is symmetric, one multiplicative Schwarz iteration corresponds to the application of a non-symmetric linear operator.
However, for an even number of iterations, the method is symmetrized by reversing the order of subdomains and, in the face-centered case, also the order of sweeps in each iteration.

\begin{figure}
\centering
\includegraphics[scale=0.8]{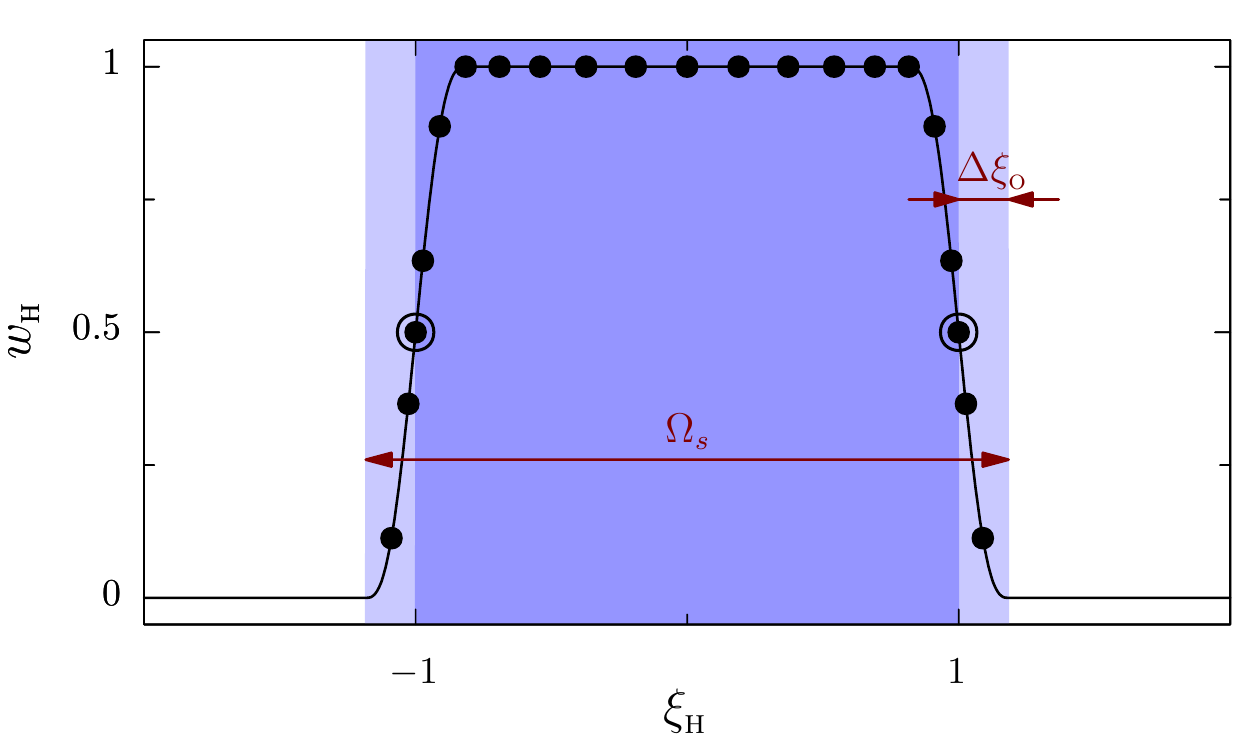}
\caption{%
Hat-shaped weight distribution $w_{\textsc h}$
for element-centered subdomains and tangential weighting in face-centered subdomains
using a quintic shape function with $P=16$ and $\no=3$.
The core region and the overlap zone of the subdomain are identified by dark and light shading, respectively.
Filled circles indicate the node positions.
On each side, the overlap zone includes three nodes from the adjoining element.
The enclosing circles mark the positions where one of these coincides with a node of the core element.
}
\label{fig:w_H}
\end{figure}

The weighted additive Schwarz method determines all local corrections independently and computes the global correction as a linear combination of these results, i.e.
\begin{equation}
\label{eq:correction:global}
\Delta \NM u \simeq \sum_{s} \transpose{\NM R_{s}} (\NM w \Delta \NM u_{s}) \, ,
\end{equation}
where ${\NM w = \NM w_2 \otimes \NM w_1}$ is a diagonal local weighting matrix that is generated from generic 1D weight distributions $\NM w_d$.
For the element-centered approach we reuse the weight distributions introduced in \cite{Stiller2016jsc} for the continuous case, i.e., $\NM w_d$ is computed from the hat-shaped weighting function
\begin{equation}
\label{eq:weight function:ec}
w_{\textsc h}(\xi_{\textsc h})
= \frac{1}{2}
  \left[ \phi_i\left(\frac{1 + \xi_{\textsc h}}{\Delta\xi_{\textsc o}}\right)
       + \phi_i\left(\frac{1 - \xi_{\textsc h}}{\Delta\xi_{\textsc o}}\right)
  \right]
\, ,
\end{equation}
where
$\xi_{\textsc h}$ is the 1D standard coordinate extended beyond ${[-1,1]}$ and
$\phi_i$ a transitional function.
Let $\Omega^{m}$ denote the element associated with the core region of subdomain $\Omega_s$, $\Omega^{m-1}$ its predecessor in the weighting direction and, respectively,
$\Omega^{m+1}$ its successor.
Then $\xi_{\textsc h}$ is computed as
\begin{equation*}
\xi_{\textsc h} =
\begin{cases}
\xi        & \text{in } \Omega^{m}        \\
\xi  \pm 2 & \text{in } \Omega^{m \pm 1}
\end{cases}
\; .
\end{equation*}
For $\phi_i$ we consider the smoothed sign functions defined as
\begin{equation*}
\phi_i(x)
= \begin{cases}
    \tilde\phi_i(x) &  x \in [-1,1] \\
    \sgn(x)         &  \text{else}
  \end{cases}
\; ,
\end{equation*}
where $\tilde\phi_i$ is a polynomial of degree ${i\in\{1,3,5,\dots\}}$ satisfying the conditions
\begin{align*}
& \tilde\phi_i(\pm 1) = \pm 1
\, ,
\\
& \frac{\D^k \tilde\phi_i}{\D x^k}(\pm 1) = 0
\, , \quad
0 < k \le (i-1)/2
\, .
\end{align*}
The $\tilde\phi_i$ are strictly monotonic in $(-1,1)$ and yield a smooth transition of the weight function in the overlap zone.
In the following we use the cubic and quintic transitions
\begin{align*}
\tilde\phi_3 &= (3x - x^3) / 2
\, ,
\\
\tilde\phi_5 &= (15x - 10x^3 +  3x^5) / 8
\, .
\end{align*}
Figure~\ref{fig:w_H} exemplifies the resulting weight distribution for the quintic case.
Other possible choices include conventional additive Schwarz, ${\phi_i = 1}$, and arithmetic averaging, ${\tilde\phi_i = 0}$, as proposed by \citet{LF05} for SEM.

\begin{figure}
\centering
\includegraphics[scale=0.8]{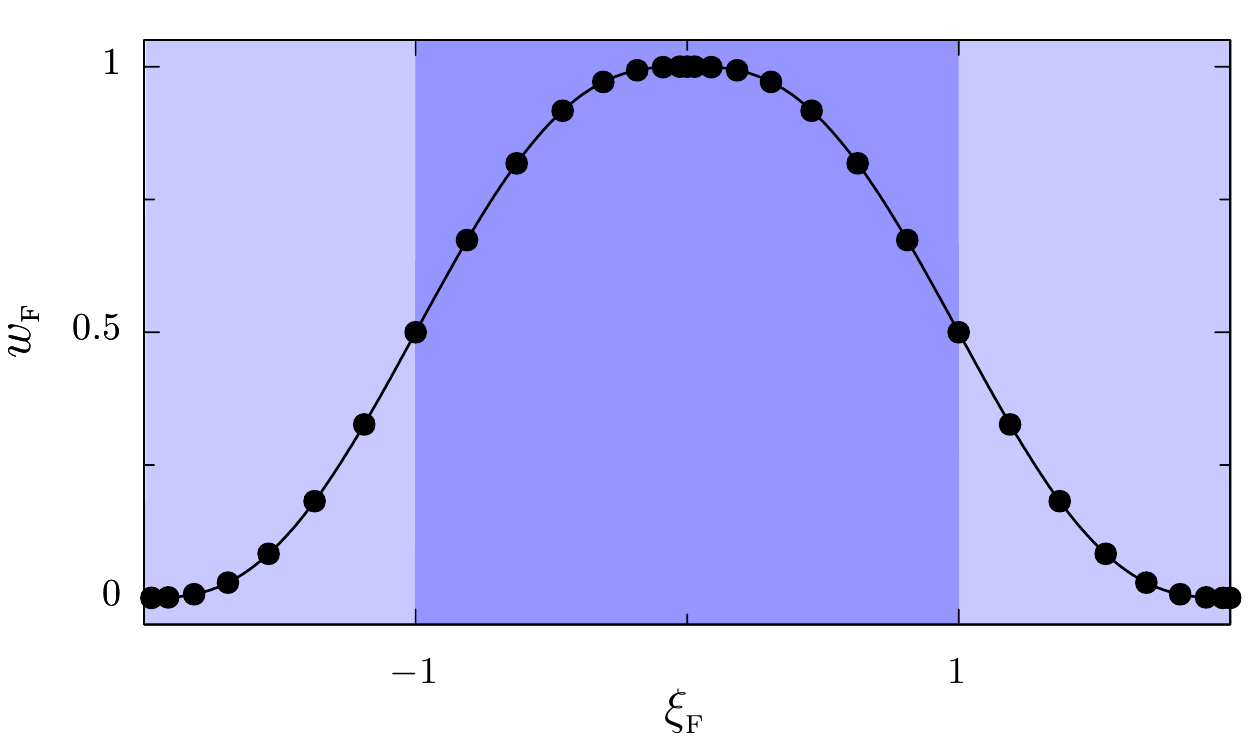}
\caption{%
Weight distribution $w_{\textsc f}$ for normal weighting in face-centered subdomains
using a quintic shape function with $P=16$. The core region and the overlap zone of
the subdomain are identified by dark and light shading, respectively.
Filled circles indicate the node positions.
}
\label{fig:w_F}
\end{figure}

With the face-centered approach, one iteration consists of one sweep over all subdomains of one orientation, followed by second sweep for the other orientation.
Note that this method is not strictly additive, since the second sweep builds on the result of the first one.
The weight matrix $\NM w$ is constructed similarly as in the element-centered case.
For the direction normal to the face consider the two adjacent elements $\Omega^{m}$ and $\Omega^{m+1}$ such that the face is located at ${\xi=1}$ in the former and at ${\xi=-1}$ in the latter. Note that we use a 1D notation for simplicity.
The normal weighting function is then defined as
\begin{equation}
\label{eq:weight function:fc}
w_{\textsc f}(\xi_{\textsc f}) =
\frac{1}{2} \left[1 + \phi_i(\vert\xi_{\textsc f}\vert) \right]
\, ,
\end{equation}
where
\begin{equation*}
\xi_{\textsc f}
= \begin{cases}
    \xi - 1 & \text{in } \Omega^{m} \\
    \xi + 1 & \text{in } \Omega^{m+1}
  \end{cases}
\; .
\end{equation*}
Figure~\ref{fig:w_F} illustrates $w_{\textsc f}$ for for the quintic case (${i=5}$).
The tangential weights are identical to those used in the element-centered case.

\subsection{Multigrid}
\label{sec:mg}

For MG we define a series of polynomial levels $\{P_l\}$
with
\mbox{$P_l = 2^l$}
increasing from $1$ at \mbox{$l=0$} to $P$ at top level $L$.
More general series can be used supposing that ${P_{l+1} > P_l}$,
but are not considered here.
Correspondingly, let
$\NM u_l$ denote the global coefficients and
$\NM A_l$ the system matrix on level $l$.
On the top level we have
\mbox{$\NM u_L = \NM u$} and
\mbox{$\NM A_L = \NM A$},
whereas on lower levels $\NM u_l$ is the defect correction and $\NM A_l$ the counterpart of $\NM A$ obtained with elements of order $P_l$.
For transferring the correction from level \mbox{$l-1$} to level $l$ we use the embedded interpolation operator $\IOP_l$, and for restricting the residual its transpose.
These ingredients allow to build a multigrid V-cycle, which is identical to the continuous case \cite{Stiller2016jsc}, but repeated for convenience in Algorithm~\ref{alg:V-cycle}.
The \textsc{Smoother} is designed as a generic procedure which employs either the multiplicative or the weighted additive Schwarz method with element-centered or face-centered subdomains.
To allow for variable V-cycles \cite{Bra93}, the number of pre- and post-smoothing steps, $\ns[1,l]$ and $\ns[2,l]$, may change from level to level.
Line~\ref{alg:V-cycle:coarse} of Algorithm~\ref{alg:V-cycle} defines the coarse grid solution by means of the pseudoinverse $\NM A_0^{+}$.
In our implementation the coarse problem is solved using the conjugate gradient method. To ensure convergence in spite of singularity, the right side is projected to the null space of $\NM A_0$, as advocated by \citet{Kaa88}.

\begin{algorithm}[ht]
\caption{Multigrid V-cycle.}
\label{alg:V-cycle}
\begin{algorithmic}[1]
\Function{MultigridCycle}{$\NM u$, $\NM f$, $\NM\ns$}
   \State $\NM u_L \gets \NM u$
   \State $\NM f_L \gets \NM f$
   \For{$l=L,1$ \textbf{step} $-1$}
       \If{$l < L$}
          \State $\NM u_l \gets 0$
       \EndIf
       \State $\NM u_l \gets$ \Call{Smoother}{$\NM u_l$, $\NM f_l$, $\ns[1,l]$}
         \Comment{Pre-smoothing}
       \State $\NM f_{l-1} \gets \transpose{\IOP}_l (\NM f_l - \NM A_l \NM u_l)$
         \Comment{Residual restriction}
   \EndFor
   \State $\NM u_0 \gets \NM A_0^{+} \NM f_0$
     \label{alg:V-cycle:coarse}
     \Comment{Coarse grid solution}
   \For{$l=1,L$}
       \State $\NM u_l \gets \NM u_l + \IOP_l \NM u_{l-1}$
         \Comment{Correction prolongation}
       \State $\NM u_l \gets$ \Call{Smoother}{$\NM u_l$, $\NM f_l$, $\ns[2,l]$}
         \Comment{Post-smoothing}
   \EndFor
   \State \textbf{return} $\NM u \gets \NM u_L$
\EndFunction
\end{algorithmic}
\end{algorithm}

\subsection{Preconditioned conjugate gradients}
\label{sec:mgcg}

Robustness and efficiency of multigrid can be enhanced by Krylov acceleration \cite{TOS00}.
Here we follow the strategy devised \cite{Stiller2016jsc}, where the inexact preconditioned conjugate gradients of \citet{GY99} were adopted to cope with the (slight) asymmetry introduced by the Schwarz method.
The resulting MGCG method is summarized in Algorithm~\ref{alg:MGCG}.
Note that, as before with the coarse problem, the right side $\NM f$ must be in the null space of $\NM A$ if the system is singular.

\begin{algorithm}[t]
\caption{Inexact multigrid-preconditioned conjugate gradients.}
\label{alg:MGCG}
\begin{algorithmic}[1]
\Function{MGCG}{$\NM u$, $\NM f$, $\NM\ns$, $i_{\max}$, $r_{\max}$}
   \State $\NM r_{\text{old}} \gets \NM 0$
   \State $\NM r \gets \NM f - \NM A \NM u$
   \State $\NM p \gets$ \Call{MultigridCycle}{$\NM 0$, $\NM r$, $\NM\ns$}
   \State $\delta \gets \transpose{\NM p} \NM r$
   \For{$i = 1, i_{\max}$}
      \State $\NM q \gets \NM A \NM p$
      \State $\alpha \gets \delta / (\transpose{\NM p} \NM q)$
      \State $\NM u \gets \NM u + \alpha \NM p$
      \State $\NM r \gets \NM r - \alpha \NM q$
      \State \textbf{if} $\lVert \NM r \rVert \le r_{\max}$ \textbf{exit}
      \State $\NM z \gets$ \Call{MultigridCycle}{$\NM 0$, $\NM r$, $\NM\ns$}
      \State $\beta \gets \transpose{\NM q}(\NM r - \NM r_{\text{old}}) / \delta$
      \State $\NM p \gets \NM z + \beta \NM p$
      \State $\delta \gets \transpose{\NM z} \NM r$
      \State $\NM r_{\text{old}} \gets \NM r$
   \EndFor
   \State \textbf{return} $\NM u$
\EndFunction
\end{algorithmic}
\end{algorithm}


\section{Results}
\label{sec:results}

For assessing robustness and efficiency, the described methods were implemented in Fortran and applied to the test case of Lottes and Fischer \cite{LF05,Fis15}, i.e.,
\begin{equation*}
-\nabla^2 u = 2\pi^2 \sin(\pi x_1) \sin(\pi x_2)
\end{equation*}
in the domain ${\Omega = (0, 2 AR)\times (0, 2)}$ with the aspect ratio
${AR \in \mathbb N}$.
Assuming periodic boundary conditions, the exact solution is
${u = \sin(\pi x_1) \sin(\pi x_2)}$
for arbitrary $AR$.
To keep the test series manageable, we constrained ourselves to equidistant grids with an identical number of elements in each direction, i.e., ${\nel[,1]=\nel[,2]}$.
As a consequence, the element aspect ratio $\Delta x_1 / \Delta x_2$ is equivalent to the domain aspect ratio $AR$.
The code was compiled using the GNU compiler collection 6.0 with optimization -O3 and executed on a 3.1\,GHz Intel Core i7-5557U CPU.
In all test runs, the initial guess was chosen at random with values confined to the interval ${[0,1]}$.


\subsection{Performance metrics}
\label{sec:performance metrics}

The primary convergence measure is the average multigrid convergence rate
\begin{equation}
\label{eq:rho}
\rho
= \sqrt[n]{\frac{r_n}{r_0}}
\, ,
\end{equation}
where $r_n$ is the Euclidean norm of the residual vector after the $n$th cycle.
Since $\rho$ varied by several orders of magnitude in some tests, we use alternatively the logarithmic convergence rate
\begin{equation}
\label{eq:r}
\bar r
= -\log_{10} \rho
\end{equation}
and the number of cycles $n_{10}$ needed to reduce the residual by a factor of $10^{10}$.
\begin{remark}
The multigrid convergence rate and, consequently, the logarithmic convergence rate depend on the number of cycles $n$.
However, this dependence is weak, except for very low $n$ or the case that machine accuracy is reached.
Both effects are negligible in the reported tests and, hence, not further considered here.
\end{remark}

As an efficiency measure we define the average equivalent number of operator applications required for reducing the residual by a factor of
$10$,
\begin{equation}
\label{eq:omega}
\bar\omega =  \bar r \frac{W_{\text{cyc}}}{W_{\text{op}}}
\, ,
\end{equation}
where
  $W_{\text{op}}$ is the cost for one application of the system matrix $\NM A$
and
  $W_{\text{cyc}}$ for one cycle.
Assuming that sum factorization is exploited, the former can be estimated as
  ${W_{\text{op}} = 2 \np^3 \nel}$.
The cycle cost comprises the contributions of Schwarz iterations, residual evaluations, the coarse grid solver and transfer operators.
Using the estimates given in Sec.~\ref{sec:Schwarz} for the first and
neglecting the latter two we arrive at
\begin{equation}
\label{eq:W_cyc}
W_{\text{cyc}}
\simeq \left[\cs\ns \left(\frac{1}{2}\cd\md + 1\right) + C_{\textsc{cg}} \right] W_{\text{op}}
\, ,
\end{equation}
where
$\ns$ is the number of pre- and post-smoothing steps on the finest level,
${\cs = \sfrac{4}{3}}$ for the classical V-cycle and $2$ for a variable V-cycle doubling the number of smoothing steps with lower levels \cite{Bra93},
$\md$ the number of sweeps per Schwarz iteration, i.e.
${\md=1}$ for element-centered and
${\md=2}$ for face-centered subdomains,
and
${C_{\textsc{cg}} = 1}$ the extra cost for conjugate gradients when using MGCG.
The ``1'' in the inner braces stems from the residual evaluation, which is also the dominant cost with CG.

Occasionally, the number of operations to achieve a certain residual reduction may be of interest.
As a representative measure we consider the average number of multiplications per unknown for a reduction by ten orders of magnitude, which can be estimated as
\begin{equation}
\label{eq:w_10}
w_{10}
\simeq
  \frac{10}{\bar r} \frac{W_{\text{cyc}}}{\np^2 \nel}
= \frac{20}{\bar r}\left[\cs\ns \left(\frac{1}{2}\cd\md + 1\right) + C_{\textsc{cg}} \right]  \np
\, .
\end{equation}


\subsection{Qualitative behavior of Schwarz methods}
\label{sec:results:qualitative}

\begin{figure}
\subfloat[Initial error.]
  {\includegraphics[width=0.4999\textwidth]{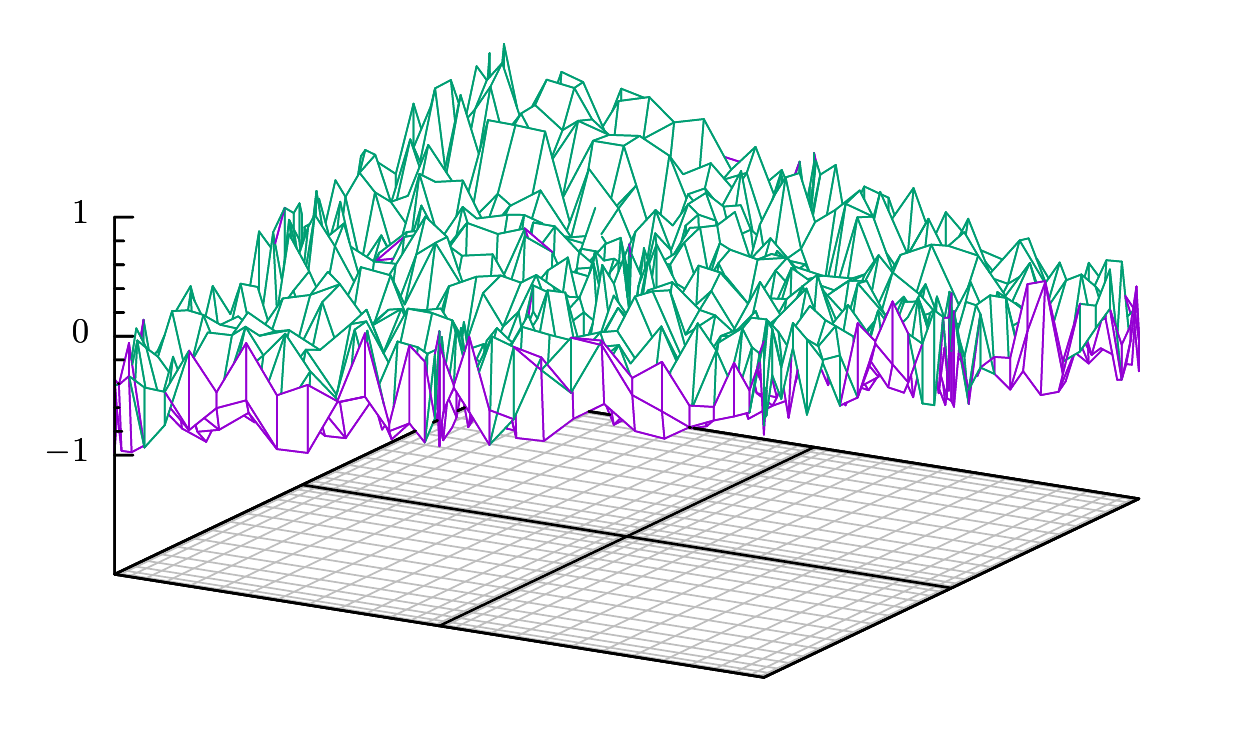}
   \label{fig:schwarz:error:initial}}
\subfloat[Element-centered additive, ${\no=0}$.]
  {\includegraphics[width=0.4999\textwidth]{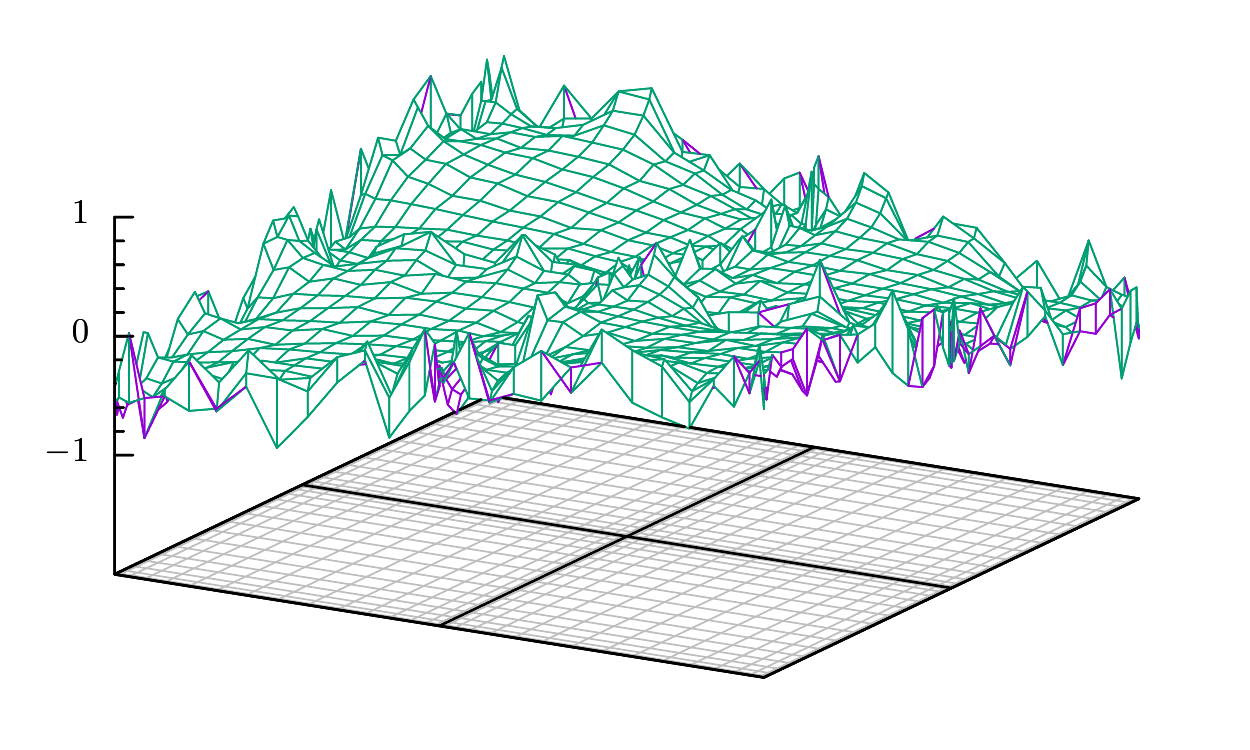}
   \label{fig:schwarz:error:ec_o0}}
\\
\subfloat[Element-centered additive, ${\no\!=\!3}$, unweighted]
  {\includegraphics[width=0.4999\textwidth]{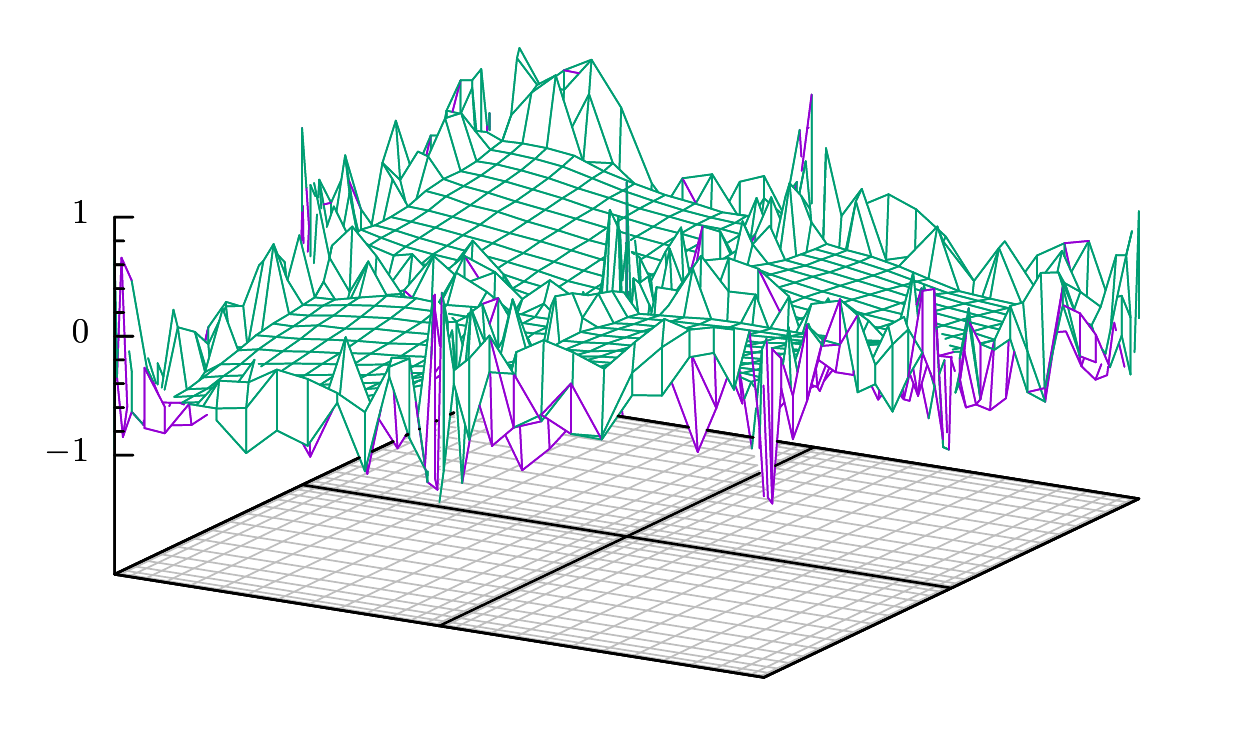}
   \label{fig:schwarz:error:ec_o2_additive}}
\subfloat[Element-centered additive, $\no=3$, arithmetic averaging.]
  {\includegraphics[width=0.4999\textwidth]{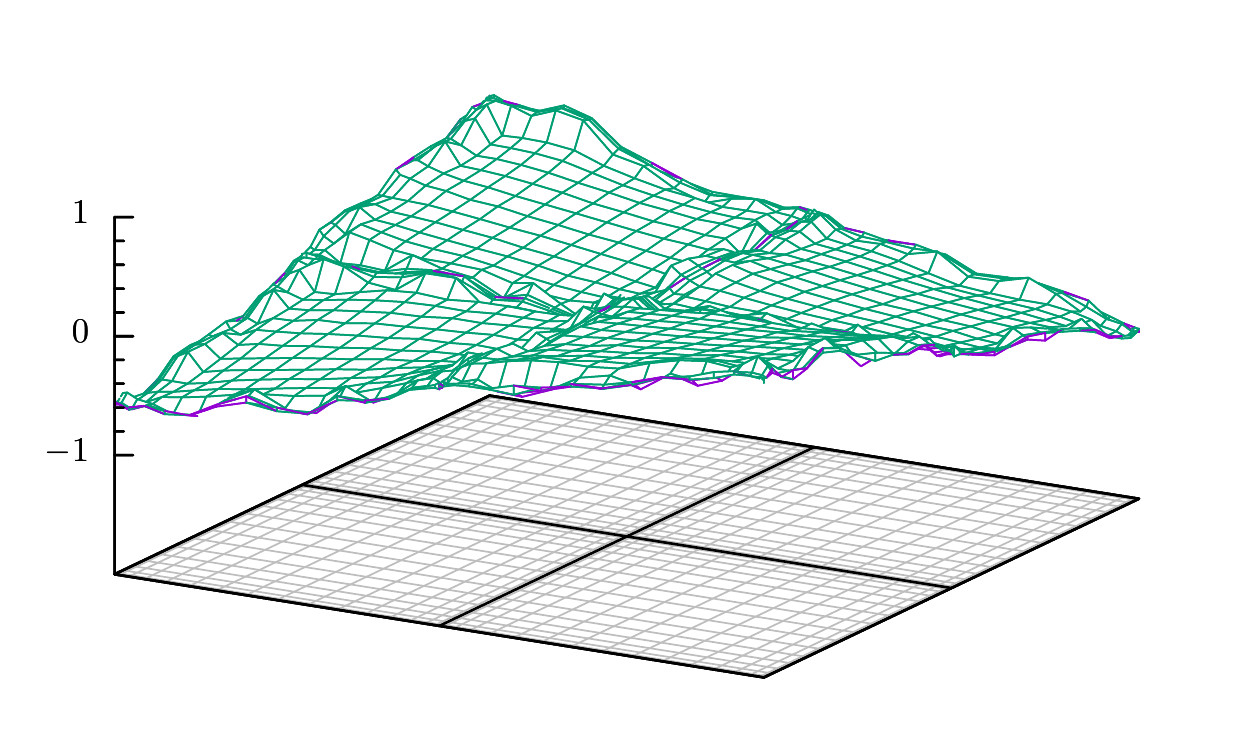}
   \label{fig:schwarz:error:ec_o2_arithmetic}}
\\
\subfloat[Element-centered additive, $\no=3$, cubic weighting.]
  {\includegraphics[width=0.4999\textwidth]{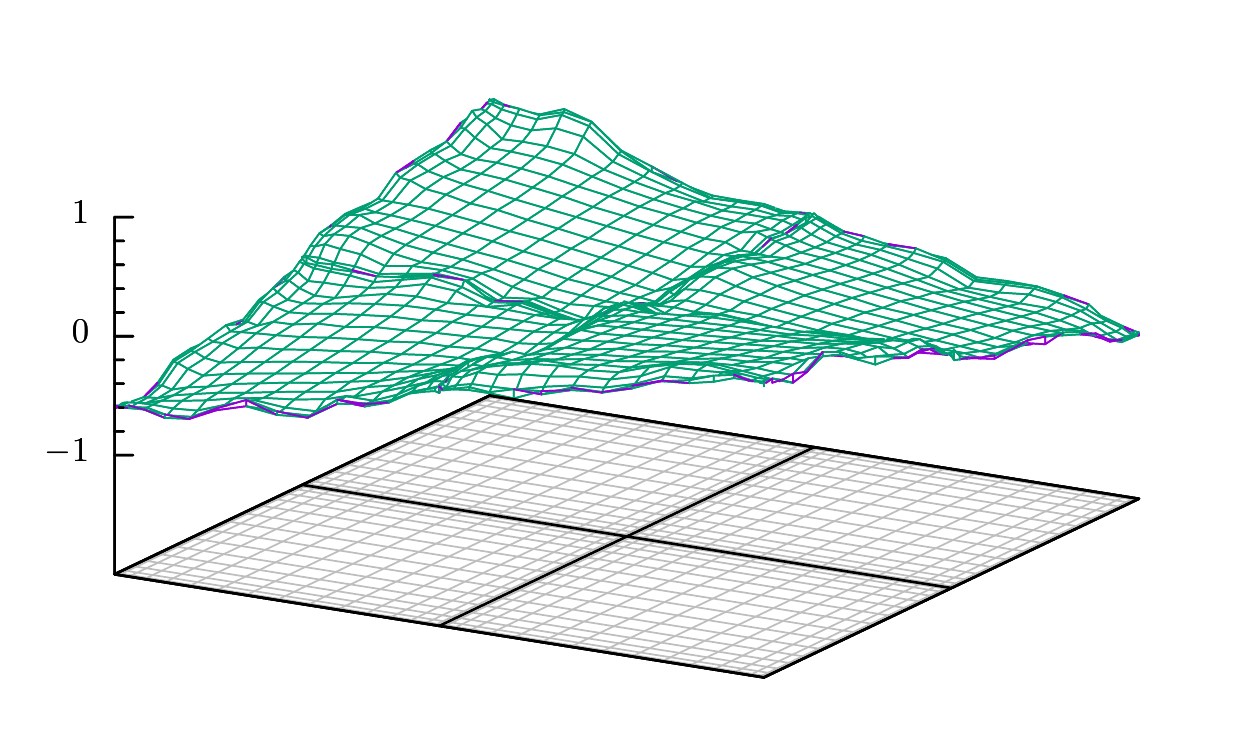}
   \label{fig:schwarz:error:ec_o2_cubic}}
\subfloat[Face-centered additive, $\no=3$, cubic weighting.]
  {\includegraphics[width=0.4999\textwidth]{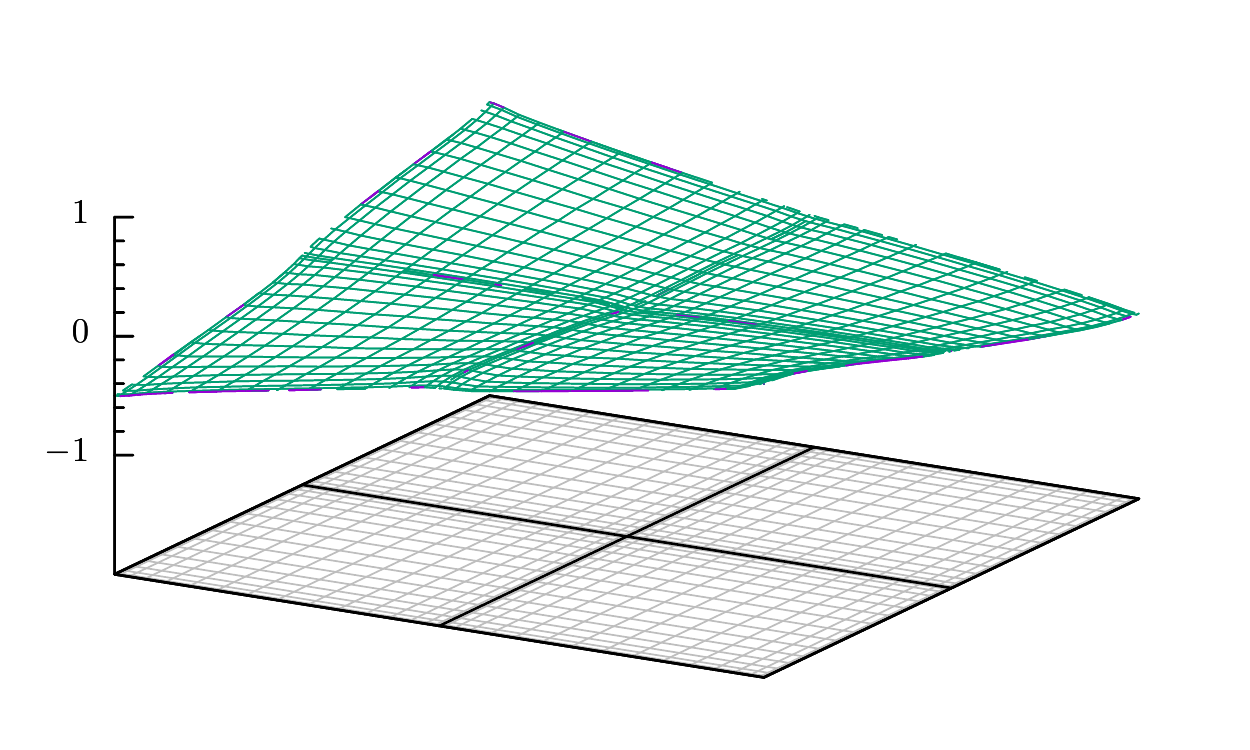}
   \label{fig:schwarz:error:fc_o2_cubic}}
\caption{Smoothing properties of selected Schwarz methods. DG with ${8\times8}$
elements of order ${P=16}$, ${\mu_{\star}=1}$ and ${\MS\beta=0}$.
Graph (a) shows the initial error in  a subregion consisting of four
elements, and (b--f) the error after one Schwarz iteration.
\label{fig:schwarz:error}
}
\end{figure}

Figure~\ref{fig:schwarz:error} illustrates the smoothing properties of selected additive Schwarz methods for a discretization using ${8\times8}$ elements of order ${P=16}$ with stabilization ${\mu_{\star}=1}$ and auxiliary parameter ${\MS\beta=0}$.
The displayed error is defined as the difference between the approximate and exact solutions adjusted to a zero median.
For clarity, the plots were restricted to a subregion comprising four elements.
Figure~\ref{fig:schwarz:error:initial} depicts the error of the random initial guess and
Figs.~\ref{fig:schwarz:error:ec_o0}-\subref*{fig:schwarz:error:fc_o2_cubic} the error after one iteration with different Schwarz methods.
In particular, Fig.~\ref{fig:schwarz:error:ec_o0} and Fig.~\ref{fig:schwarz:error:ec_o2_additive} reveal that the non-overlapping element-centered method and the unweighted element-centered method with overlap ${\no=3}$ fail to smooth the error across the element boundaries.
The jumps produced with both methods tend to dominate the residual and lead to a severe degradation of MG efficiency.
Arithmetically weighting the overlapping Schwarz updates greatly improves this behavior, although the error still exhibits ridges near the element boundaries (Fig.~\ref{fig:schwarz:error:ec_o2_arithmetic}).
Using a smooth hat-shaped weight distribution removes these ridges and yields the best smoothing properties for element-centered subdomains (Fig.~\ref{fig:schwarz:error:ec_o2_cubic}).
Finally, Figure~\ref{fig:schwarz:error:fc_o2_cubic} illustrates the excellent performance of the overlapping face-centered Schwarz method with cubic weighting.
It should be noted, however, that one face-centered iteration has two sweeps instead of one with element-centered domains and, in addition to this, employs a larger overlap into the face normal direction.

\subsection{Convergence and robustness}
\label{sec:results:convergence}
\label{sec:results:robustness}

To investigate the influence of overlap and weighting, and to assess the robustness of the multigrid approach we performed several numerical experiments on grids consisting of up to $256^2$ elements of order ${P=4}$ to 32 and aspect ratios between 1 and 32.
For concise notation we use two-letter acronyms, where the first letter identifies the subdomain type  (``E'' -- element-centered, ``F'' -- face-centered) and the second the iteration method (``M'' -- multiplicative, ``A'' additive).
Additionally, subscript ``0'' indicates zero overlap, ${\no=0}$, and ``$\ell$'' a level-dependent overlap of ${\no[,l] = 1\!+\!\lfloor P_l/8 \rfloor}$.
For example, FM\tsub{0} denotes the face-centered multiplicative Schwarz method with ${\no=0}$, i.e. no lateral overlap.
If not indicated otherwise, one pre-smoothing and one post-smoothing step were applied on each level ${l>0}$.
The weighting method (cubic or quintic) is stated where necessary.

Table~\ref{tab:overlap,weighting} shows the convergence rates of selected MG and MGCG methods
on a uniform ${16\times16}$ tessellation of the domain ${[0,2]^2}$.
The results for the element-centered smoothers resemble those obtained with continuous spectral elements \cite{Stiller2016jsc}.
Using the multiplicative smoother with no overlap, EM\tsub{0},
MG reaches convergence rates up to  ${\bar r = 0.63}$ with ${P=4}$
and MGCG up to ${\bar r = 0.9}$, but both degrade with growing polynomial order $P$.
In the additive case, EA\tsub{0}, MG fails to converge (not shown), whereas MGCG just succeeds.
Lines 7\textminus14 show the results obtained with the level-dependent overlap, ${\no[,l] = 1\!+\!\lfloor P_l/8 \rfloor}$.
It imposes a lower bound on the overlap width $\deltaO$ that corresponds approximately to one eighth of the neighbor element width in the direction normal to the boundary.
With this choice MG becomes robust against increasing $P$ for multiplicative as well as additive Schwarz smoothing.
Comparing the smoothers reveals that EA\tsub{\ell} consistently achieved a higher convergence rates than EM\tsub{\ell}.
As a possible reason we found that the instantaneous updates in the multiplicative method tend to produce excessive gradients in overlap regions,
whereas the additive method avoids this peculiarity by applying a weighted average.
Both, cubic as well as quintic weighting are suited, though the latter proved
slightly more efficient.

Lines 15\textminus28 of Tab.~\ref{tab:overlap,weighting} show the results for the face-centered smoothers.
Compared to the element-centered smoothers, they achieve notably higher convergence rates, which can be attributed to the inbuilt overlap normal to the face and the double sweep over both coordinate directions.
Even with no lateral overlap the approach proved robust against $P$.
Application of a level-dependent lateral overlap, ${\no[,l] = 1\!+\!\lfloor P_l/8 \rfloor}$, increased the logarithmic convergence rate to magnitudes in the range between 2 and 3.5.
In contrast to the element-centered method, the difference between cubic and quintic weighting is marginal and hence not considered here.
Moreover, the face-centered method exhibits a lower sensitivity with respect to the auxiliary LDG parameter $\MS\beta$.
In most cases, similar convergence rates were obtained for central and non-central numerical fluxes, i.e.\ ${\MS\beta=0}$ and ${\MS\beta=\sfrac{1}{2}}$, whereas the latter tends to converge slower than the former when using element-centered smoothers.

In the following we focus our attention to MGCG for IP/LDG with central numerical fluxes and abandon cubic in favor of quintic weighting.
Table~\ref{tab:robustness:size} shows the logarithmic convergence rates $\bar r$, cycle counts $n_{10}$ and equivalent operator applications $\bar\omega$ for smoothers
EM\tsub{0},
EA\tsub{\ell},
FA\tsub{0} and
FA\tsub{\ell}
on equidistant grids ranging from ${8^2}$ to ${256^2}$ elements and polynomial orders from ${P=4}$ to 32.
The results clearly confirm that the methods are robust with respect to the grid size.
As expected, EM\tsub{0} degrades for growing $P$, whereas with EA\tsub{\ell}, FA\tsub{0} and FA\tsub{\ell} the increased order leads to even higher convergence rates and, hence, lower cycle counts.
Comparing the average number of operator applications required for reducing the residual by a factor of 10, we find that EA\tsub{\ell} is the most efficient method with ${\bar\omega\in[6,11]}$.
For example, consider order ${P=16}$ with ${\bar r \approx 2.2}$:
taking into account
${\cs = 4/3}$,
${\ns = 2}$,
${\cd = 4(1 + 2\co)^3}$,
${\co \lesssim 3/17}$,
${C_{\textsc{cg}} = 1}$ and
${\md = 1}$,
the estimate \eqref{eq:w_10} results in
just 2608 multiplications per unknown to achieve a residual reduction by ten orders of magnitude.
EM\tsub{0} uses no overlap and, hence attains lower cycle costs.
As a result it remains competitive with EA\tsub{\ell} for ${P=4}$ and still the second-best method for ${P=8}$, despite the higher cycle count.
FA\tsub{\ell} converges about 1.4 times faster than EA\tsub{\ell}, but fails to compensate the sixfold higher operation count and thus remains about twice as expensive in terms of $\bar\omega$.

\begin{figure}
\centering
\subfloat[Convergence rates.]{\includegraphics[width=0.8\textwidth]{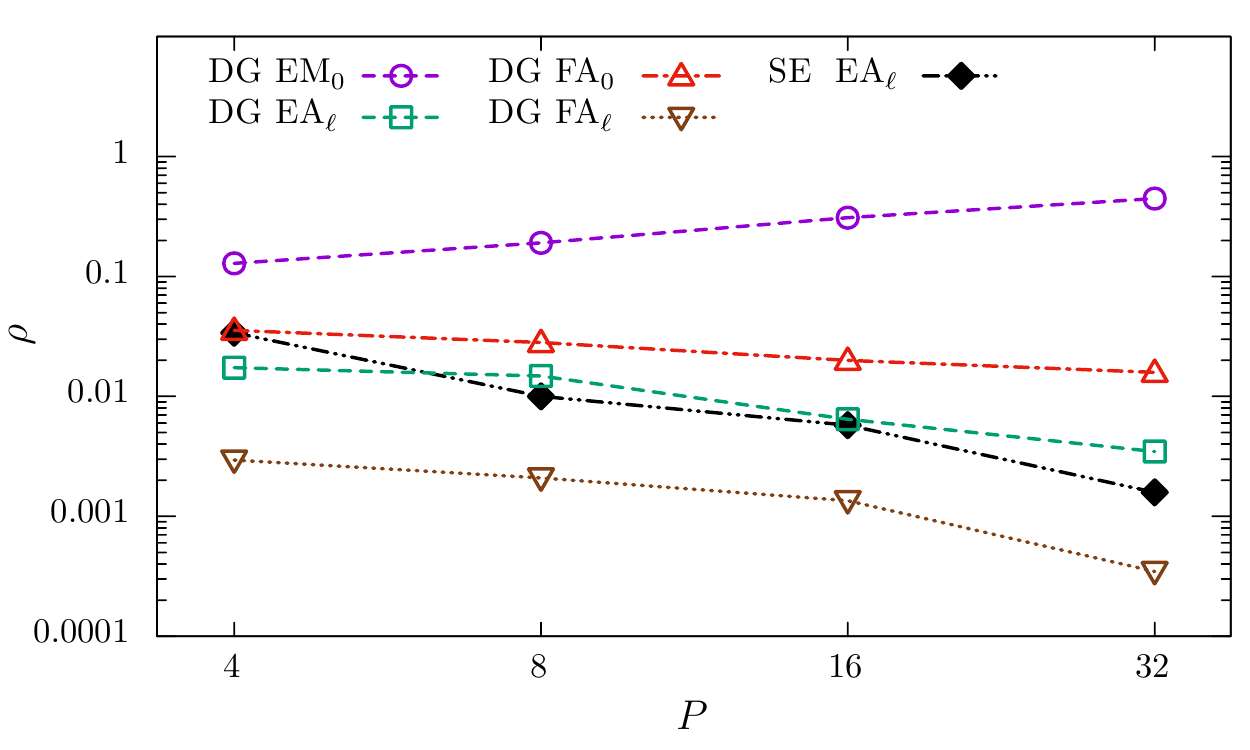}
\label{fig:p:rho}}
\\
\subfloat[Runtimes.]{\includegraphics[width=0.8\textwidth]{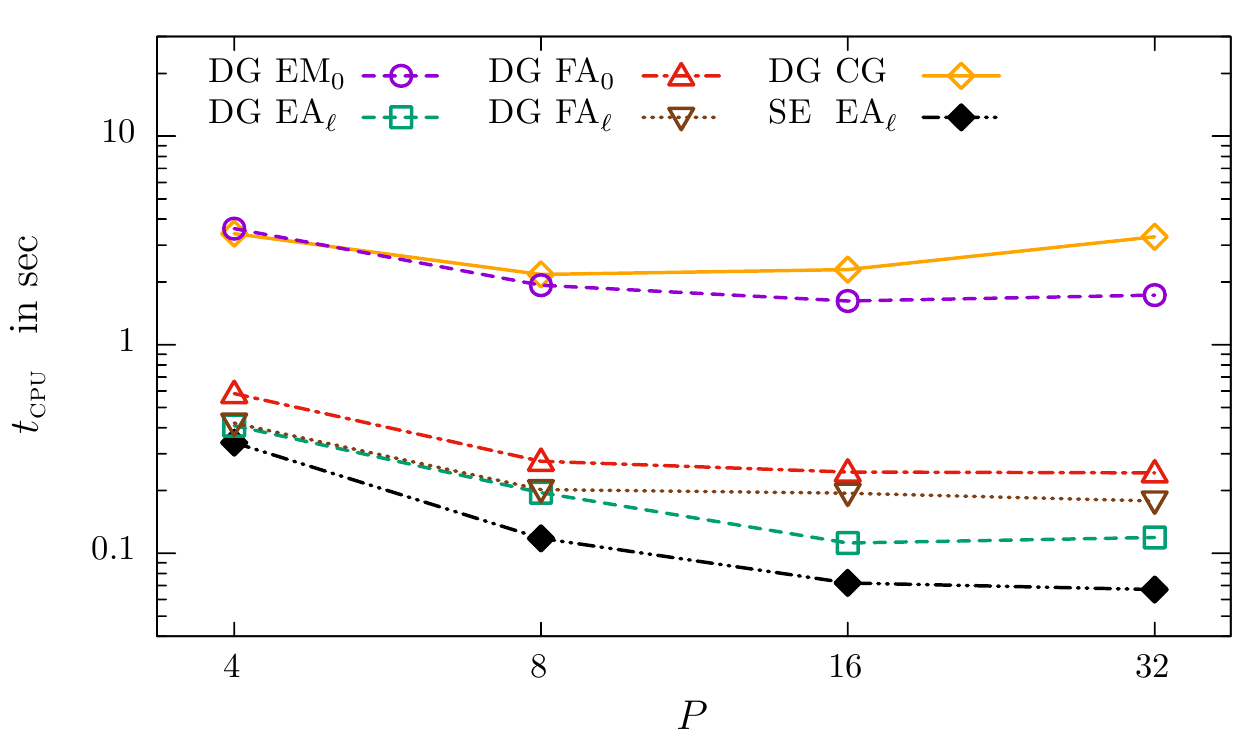}
\label{fig:p:t}}
\caption{MGCG convergence rates and runtimes for a $10^{10}$ residual reduction using $(256/P)^2$ square elements of order $P$. DG denotes the discontinuous Galerkin method with parameters and smoothers according to Tab.~\ref{tab:robustness:size},
SE the corresponding spectral element method, and
CG the DG conjugate gradient solver.
\label{fig:p}
}
\end{figure}

Figure~\ref{fig:p:rho} shows the multigrid convergence rates in comparison with the spectral element (SE) version of EA\tsub{\ell}.
In line with our expectations the comparison asserts that the discontinuous and continuous methods converge nearly with identical rates, improving from ${\rho \approx 0.02}$ at ${P=4}$ to 0.003 at ${P=32}$.
The face-centered additive smoother yields even faster convergence with rates between 0.003 and 0.0003.
To assess the actual computational cost, Figure~\ref{fig:p:t} depicts the runtimes required to reduce the residual by ten orders of magnitude for different polynomial degrees.
Note that the number of elements was adjusted according to ${\nel[,d]=\sfrac{256}{P}}$ to assure a nearly constant problem size.
Sum factorization was exploited on levels ${P_l\ge8}$.
Except for the CG method, which included for comparison, all graphs exhibit a gentle downward slope which becomes increasingly horizontal with growing $P$.
This indicates that the increased operation count scaling as $\bar\omega P$ is more than compensated by the gain in computational efficiency due to the larger operator size.
For the same reason, the face-centered methods, FA\tsub{0} and FA\tsub{\ell} come considerably closer to EA\tsub{\ell} than predicted by $\bar\omega$.
In contrast, EM\tsub{0} performs much worse than expected.
It consumes more than 10 times the runtime of EA\tsub{\ell} and barely outperforms CG.
This behavior is a consequence of the recursive solution strategy of multiplicative Schwarz: It prevents the ``stacking'' of operands as in the additive case, where it allows to convert matrix-vector multiplications into vastly more efficient matrix-matrix multiplications.
The fastest discontinuous method, MGCG with EA\tsub{\ell}, solves the problem with ${P=16}$ in 0.112\,s or, respectively, 1.5\,\textmu s per unknown.
Counting only multiplications, this corresponds to a performance of 1.9\,GFLOPS.
The spectral element counterpart attains nearly identical convergence rates, but succeeds with just two thirds of the runtime, which corresponds almost exactly to the lower operation count.

\begin{table}
\centering
\caption{Convergence rates of MG and MGCG with different overlaps and weighting methods. DG with ${\mu_{\star}=1}$ and constant $\MS\beta$ using a uniform 16\,\texttimes\,16 grid.}
\label{tab:overlap,weighting}
\centering
\begin{tabular}{cccccccccccc}
\toprule
  &       &      &         &       & \multicolumn{4}{c}{$\bar r$} \\
                                     \cmidrule(rl){6-9}
\# & method & smoother        & weighting & $\MS\beta$
                                             & \pbox{4}  & \pbox{8}  & \pbox{16} & \pbox{32} \\
\midrule
 1 & MG     & EM\tsub{0}      &     --    & 0     &   0.63    &   0.36    &   0.22    &   0.15    \\
 2 &        &                 &     --    & \half &   0.43    &   0.26    &   0.17    &   0.13    \\
 3 & MGCG   & EM\tsub{0}      &     --    & 0     &   0.90    &   0.72    &   0.52    &   0.36    \\
 4 &        &                 &     --    & \half &   0.73    &   0.58    &   0.40    &   0.28    \\
 5 &        & EA\tsub{0}      &     --    & 0     &   0.20    &   0.09    &   0.03    &   0.01    \\
 6 &        &                 &     --    & \half &   0.16    &   0.06    &   0.02    &   0.01    \\
\midrule
 7 & MG     & EM\tsub{\ell} &     --    & 0     &   1.02    &   1.01    &   1.13    &   1.45    \\
 8 &        &                 &     --    & \half &   0.61    &   0.84    &   0.92    &   1.16    \\
 9 &        & EA\tsub{\ell} &  cubic    & 0     &   1.39    &   1.64    &   1.82    &   1.99    \\
10 &        &                 &  cubic    & \half &   1.52    &   1.69    &   1.70    &   1.98    \\
11 &        &                 &  quintic  & 0     &   1.66    &   1.65    &   2.11    &   2.51    \\
12 &        &                 &  quintic  & \half &   1.56    &   1.68    &   2.04    &   2.49    \\
13 & MGCG   & EA\tsub{\ell} &  quintic  & 0     &   1.76    &   1.84    &   2.20    &   2.49    \\
14 &        &                 &  quintic  & \half &   1.60    &   1.74    &   2.07    &   2.40    \\
\midrule
15 & MG     & FM\tsub{0}      &     --    & 0     &   1.64    &   1.71    &   1.87    &   1.96    \\
16 &        &                 &     --    & \half &   1.45    &   1.34    &   1.32    &   1.34    \\
17 &        & FA\tsub{0}      &  quintic  & 0     &   1.15    &   1.22    &   1.32    &   1.37    \\
18 &        &                 &  quintic  & \half &   1.20    &   1.14    &   1.13    &   1.16    \\
19 & MGCG   & FM\tsub{0}      &     --    & 0     &   1.93    &   2.03    &   2.28    &   2.41    \\
20 &        &                 &     --    & \half &   1.65    &   1.66    &   1.72    &   1.84    \\
21 &        & FA\tsub{0}      &  quintic  & 0     &   1.45    &   1.57    &   1.70    &   1.82    \\
22 &        &                 &  quintic  & \half &   1.43    &   1.54    &   1.61    &   1.67    \\
\midrule
23 & MG     & FM\tsub{\ell} &     --    & 0     &   2.41    &   2.53    &   2.66    &   2.83    \\
24 &        &                 &     --    & \half &   2.10    &   2.54    &   3.01    &   3.18    \\
25 &        & FA\tsub{\ell} &  quintic  & 0     &   2.02    &   2.35    &   2.56    &   3.11    \\
26 &        &                 &  quintic  & \half &   2.47    &   2.61    &   3.26    &   3.53    \\
27 & MGCG   & FA\tsub{\ell} &  quintic  & 0     &   2.54    &   2.71    &   3.10    &   3.50    \\
28 &        &                 &  quintic  & \half &   2.51    &   2.62    &   3.19    &   3.30    \\
\bottomrule
\end{tabular}
\end{table}

\begin{figure}
\centering
\includegraphics[width=0.8\textwidth]{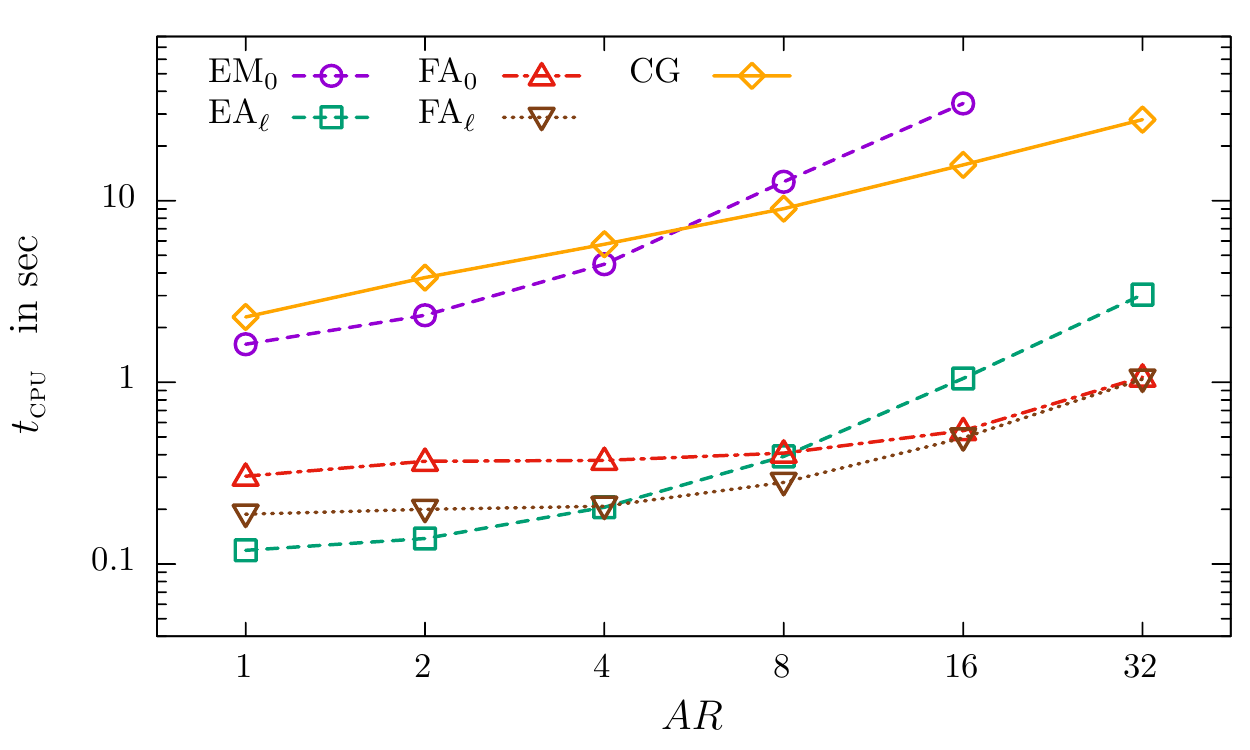}
\caption{MGCG and CG runtimes for a $10^{10}$ residual reduction at different aspect ratios. Discretization using ${16\!\times\!16}$ elements of order ${P\!=\!16}$ with 73,984 unknowns; parameters and smoothers as in Tab.~\ref{tab:robustness:AR}.
\label{fig:t:ar}}
\end{figure}

For assessing the sensitivity to the element aspect ratio,
$AR$ was varied from 1 to 32.
The domains were decomposed into ${16\times16}$ rectangular elements of the order $P$, which ranged from 4 to 32.
In addition to the methods considered above, the tests included variants using a variable V-cycle.
With the latter the number of smoothing steps is doubled when switching to the next coarser level, i.e.\ ${\ns[,l]=(2^{L-l},2^{L-l})}$.
The variable V-cycle improves convergence speed and robustness, but also raises the cost of one cycle by approximately 50\%.
In our tests we observed runtime savings in face-centered case, whereas the extra cost prevailed in the element-centered case.
According to Tab.~\ref{tab:robustness:AR}, both element-centered methods show a strong sensitivity to the aspect ratio:
Similar to CG, EM\tsub{0} degrades severely as soon as $AR$ exceeds 4, whereas EA\tsub{\ell} retains still 30 to 40 percent of the original convergence rate with ${AR=8}$.
In comparison, the face-centered methods proved rather robust, in particular with higher ansatz order.
For example, with FA\tsub{0}, $n_{10}$ multiplies by 5 when increasing $AR$ from 1 to 16 for ${P=4}$, but only by 1.4 for $P=16$.
The overlapping face-centered method, FA\tsub{\ell} exhibits a similar behavior.
Finally, Figure~\ref{fig:t:ar} shows the corresponding runtimes for ${P=16}$.
EA\tsub{\ell} remains the fastest method for ${AR \le 2}$, whereas EA\tsub{0} is by far the slowest.
Due to their better robustness, FA\tsub{\ell} and FA\tsub{0}
close up with growing aspect ratio.
They break even with EA\tsub{\ell} at ${AR=4}$ and ${AR=8}$, respectively, and gain a clear advantage for higher aspect ratios.

\begin{table}
\centering
\caption{Robustness of MGCG against the problem size; ${\mu_{\star}=1}$, ${\MS\beta = 0}$, tessellation with ${\nel[,d]\times\nel[,d]}$ square elements.
}
\label{tab:robustness:size}
\centering
\begin{tabular}{rrcccccccccccccccc}
\toprule
     &   && \multicolumn{3}{c}{EM\tsub{0}}
         && \multicolumn{3}{c}{EA\tsub{\ell}}
         && \multicolumn{3}{c}{FA\tsub{0}}
         && \multicolumn{3}{c}{FA\tsub{\ell}}
\\
\cmidrule(rl){ 4- 6}
\cmidrule(rl){ 8-10}
\cmidrule(rl){12-14}
\cmidrule(rl){16-18}
 $P$ & $\nel[,d]$ && \makebox[1.5em][c]{$\bar r$}
                   & \makebox[1.5em][c]{$n_{10}$}
                   & \makebox[1.5em][c]{$\bar{\omega}$}
                  && \makebox[1.5em][c]{$\bar r$}
                   & \makebox[1.5em][c]{$n_{10}$}
                   & \makebox[1.5em][c]{$\bar{\omega}$}
                  && \makebox[1.5em][c]{$\bar r$}
                   & \makebox[1.5em][c]{$n_{10}$}
                   & \makebox[1.5em][c]{$\bar{\omega}$}
                  && \makebox[1.5em][c]{$\bar r$}
                   & \makebox[1.5em][c]{$n_{10}$}
                   & \makebox[1.5em][c]{$\bar{\omega}$}
\\
\midrule
  4  &    8  &&  0.92  &  11  &  9.8  &&  1.78  &   6  & 10.3  &&  1.45  &   7  & 24.6  &&  2.53  &   4  & 21.5  \\
     &   16  &&  0.90  &  12  & 10.0  &&  1.76  &   6  & 10.4  &&  1.45  &   7  & 24.6  &&  2.54  &   4  & 21.4  \\
     &   32  &&  0.89  &  12  & 10.1  &&  1.76  &   6  & 10.4  &&  1.45  &   7  & 24.6  &&  2.53  &   4  & 21.5  \\
     &   64  &&  0.89  &  12  & 10.1  &&  1.76  &   6  & 10.4  &&  1.45  &   7  & 24.6  &&  2.53  &   4  & 21.5  \\
     &  128  &&  0.89  &  12  & 10.1  &&  1.76  &   6  & 10.4  &&  1.45  &   7  & 24.6  &&  2.54  &   4  & 21.4  \\
     &  256  &&  0.89  &  12  & 10.1  &&  1.76  &   6  & 10.4  &&  1.45  &   7  & 24.6  &&  2.53  &   4  & 21.5  \\
\midrule
  8  &    8  &&  0.73  &  14  & 12.3  &&  1.85  &   6  & 10.7  &&  1.55  &   7  & 23.0  &&  2.61  &   4  & 21.7  \\
     &   16  &&  0.72  &  14  & 12.5  &&  1.84  &   6  & 10.7  &&  1.57  &   7  & 22.7  &&  2.71  &   4  & 20.9  \\
     &   32  &&  0.72  &  14  & 12.5  &&  1.84  &   6  & 10.7  &&  1.57  &   7  & 22.7  &&  2.63  &   4  & 21.6  \\
     &   64  &&  0.72  &  14  & 12.5  &&  1.84  &   6  & 10.7  &&  1.57  &   7  & 22.7  &&  2.68  &   4  & 21.2  \\
     &  128  &&  0.72  &  14  & 12.5  &&  1.84  &   6  & 10.7  &&  1.57  &   7  & 22.7  &&  2.68  &   4  & 21.2  \\
     &  256  &&  0.72  &  14  & 12.5  &&  1.84  &   6  & 10.7  &&  1.57  &   7  & 22.7  &&  2.68  &   4  & 21.2  \\
\midrule
 16  &    8  &&  0.52  &  20  & 17.3  &&  2.26  &   5  &  7.5  &&  1.67  &   6  & 21.4  &&  3.15  &   4  & 16.5  \\
     &   16  &&  0.52  &  20  & 17.3  &&  2.20  &   5  &  7.7  &&  1.70  &   6  & 21.0  &&  3.10  &   4  & 16.8  \\
     &   32  &&  0.52  &  20  & 17.3  &&  2.19  &   5  &  7.7  &&  1.70  &   6  & 21.0  &&  3.17  &   4  & 16.4  \\
     &   64  &&  0.52  &  20  & 17.3  &&  2.19  &   5  &  7.7  &&  1.70  &   6  & 21.0  &&  3.11  &   4  & 16.7  \\
     &  128  &&  0.52  &  20  & 17.3  &&  2.19  &   5  &  7.7  &&  1.70  &   6  & 21.0  &&  3.11  &   4  & 16.7  \\
     &  256  &&  0.52  &  20  & 17.3  &&  2.19  &   5  &  7.7  &&  1.70  &   6  & 21.0  &&  3.12  &   4  & 16.7  \\
\midrule
 32  &    8  &&  0.36  &  28  & 25.0  &&  2.46  &   5  &  6.3  &&  1.77  &   6  & 20.2  &&  3.47  &   3  & 14.3  \\
     &   16  &&  0.36  &  29  & 25.0  &&  2.49  &   5  &  6.2  &&  1.82  &   6  & 19.6  &&  3.50  &   3  & 14.2  \\
     &   32  &&  0.36  &  28  & 25.0  &&  2.47  &   5  &  6.3  &&  1.82  &   6  & 19.6  &&  3.46  &   3  & 14.3  \\
     &   64  &&  0.36  &  28  & 25.0  &&  2.46  &   5  &  6.3  &&  1.82  &   6  & 19.6  &&  3.38  &   3  & 14.7  \\
     &  128  &&  0.36  &  28  & 25.0  &&  2.46  &   5  &  6.3  &&  1.82  &   6  & 19.6  &&  3.52  &   3  & 14.1  \\
     &  256  &&  0.36  &  28  & 25.0  &&  2.46  &   5  &  6.3  &&  1.82  &   6  & 19.6  &&  3.53  &   3  & 14.0  \\
\bottomrule
\end{tabular}
\end{table}



\begin{table}
\centering
\caption{Robustness of MGCG against the aspect ratio ${AR = \sfrac{\Delta x_1}{\Delta x_2}}$; ${\mu_{\star}=1}$, ${\MS\beta = 0}$;
EM\tsub{0}, EA\tsub{\ell} using one pre- and one post-smoothing and
FA\tsub{0}, FA\tsub{\ell} a variable V-cycle with ${\ns[,l]=(2^{L-l},2^{L-l})}$.
}
\label{tab:robustness:AR}
\centering
\begin{tabular}{rrcccccccccccc}
\toprule
     &   && \multicolumn{2}{c}{EM\tsub{0}}
         && \multicolumn{2}{c}{EA\tsub{\ell}}
         && \multicolumn{2}{c}{FA\tsub{0}}
         && \multicolumn{2}{c}{FA\tsub{\ell}}
\\
\cmidrule(rl){ 4- 5}
\cmidrule(rl){ 7- 8}
\cmidrule(rl){10-11}
\cmidrule(rl){13-14}
 $P$ & $AR$  && \makebox[1.5em][c]{$\bar r$}
              & \makebox[1.5em][c]{$n_{10}$}
             && \makebox[1.5em][c]{$\bar r$}
              & \makebox[1.5em][c]{$n_{10}$}
             && \makebox[1.5em][c]{$\bar r$}
              & \makebox[1.5em][c]{$n_{10}$}
             && \makebox[1.5em][c]{$\bar r$}
              & \makebox[1.5em][c]{$n_{10}$}
\\
\midrule
  4  &   1 &&  0.90  &   12  &&  1.76  &    6  &&  1.52  &   7 &&  2.78  &   4    \\
     &   2 &&  0.74  &   14  &&  1.26  &    8  &&  1.33  &   8 &&  2.49  &   5    \\
     &   4 &&  0.32  &   32  &&  0.88  &   12  &&  1.18  &   9 &&  1.86  &   6    \\
     &   8 &&  0.13  &   80  &&  0.47  &   22  &&  0.85  &  12 &&  1.05  &  10    \\
     &  16 &&  0.08  &  120  &&  0.04  &  236  &&  0.30  &  34 &&  0.41  &  25    \\
     &  32 &&  0.07  &  140  &&  0.03  &  321  &&  0.13  &  79 &&  0.16  &  62    \\
\midrule
  8  &   1 &&  0.72  &   14  &&  1.84  &    6  &&  1.63  &   7 &&  3.10  &   4    \\
     &   2 &&  0.56  &   18  &&  1.76  &    6  &&  1.49  &   7 &&  3.38  &   3    \\
     &   4 &&  0.29  &   35  &&  1.20  &    9  &&  1.43  &   7 &&  2.63  &   4    \\
     &   8 &&  0.12  &   87  &&  0.70  &   15  &&  1.18  &   9 &&  1.57  &   7    \\
     &  16 &&  0.07  &  141  &&  0.25  &   40  &&  0.77  &  14 &&  0.91  &  12    \\
     &  32 &&  0.06  &  178  &&  0.10  &   98  &&  0.30  &  34 &&  0.36  &  28    \\
\midrule
 16  &   1 &&  0.52  &   20  &&  2.20  &    5  &&  1.78  &   6 &&  3.63  &   3    \\
     &   2 &&  0.37  &   28  &&  2.07  &    5  &&  1.62  &   7 &&  3.64  &   3    \\
     &   4 &&  0.21  &   48  &&  1.43  &    7  &&  1.58  &   7 &&  3.33  &   3    \\
     &   8 &&  0.10  &   97  &&  0.85  &   12  &&  1.57  &   7 &&  2.58  &   4    \\
     &  16 &&  0.07  &  137  &&  0.34  &   30  &&  1.19  &   9 &&  1.53  &   7    \\
     &  32 &&  0.06  &  161  &&  0.13  &   76  &&  0.60  &  17 &&  0.80  &  13    \\
\midrule
 32  &   1 &&  0.35  &   29  &&  2.49  &    5  &&  1.89  &   6 &&  3.96  &   3    \\
     &   2 &&  0.23  &   44  &&  2.39  &    5  &&  1.78  &   6 &&  4.05  &   3    \\
     &   4 &&  0.15  &   65  &&  1.71  &    6  &&  1.80  &   6 &&  4.22  &   3    \\
     &   8 &&  0.09  &  116  &&  1.07  &   10  &&  1.80  &   6 &&  4.55  &   3    \\
     &  16 &&  0.07  &  150  &&  0.41  &   25  &&  1.64  &   7 &&  2.55  &   4    \\
     &  32 &&  0.06  &  157  &&  0.17  &   61  &&  1.07  &  10 &&  1.40  &   8    \\
\bottomrule
\end{tabular}
\end{table}


\section{Conclusions}
\label{sec:conclusions}

We presented a multigrid method for nodal discontinuous Galerkin formulations of the Poisson equation on two-dimensional Cartesian grids.
The method adopts and extends techniques developed recently for the continuous spectral element method \cite{LF05,Stiller2016jsc}.
Using the nodal basis corresponding to the Gauss-Lobatto-Legendre points in conjunction with the related quadrature we derived a unified form of the discrete equations, which embodies the interior penalty method as well as the local discontinuous Galerkin method.
These equations are solved by means of polynomial multigrid with multiplicative or weighted additive Schwarz methods for smoothing and, optionally, the inexact preconditioned conjugate gradient method \cite{GY99} for acceleration.
The Schwarz methods operate on a set of overlapping rectangular subdomains, which are either element- or face-centered.
The resulting multigrid methods achieved excellent convergence rates independent from the problem size.
Using a level-dependent overlap of ${1\!+\!\lfloor P_l/8 \rfloor}$ nodes proved sufficient for robustness against the ansatz order up to ${P=32}$.
Taking advantage of tensor-product factorization and fast diagonalization techniques, the methods attain a computational complexity of ${O(PN)}$ per cycle.
In terms of runtime, the solvers actually achieve linear complexity, since the convergence rate and the computational efficiency improve with growing order.
Multigrid with conjugate gradient acceleration and the element-centered additive smoother with level-dependent overlap
is the best choice for equidistant grids with nearly square elements, for which it achieves convergence rates between ${\rho=0.017}$ and and 0.003.
With ${P=16}$ it needs about 2600 multiplications per unknown to reduce the residual by ten orders of magnitudes.
Multigrid with conjugate gradient acceleration and the face-centered additive smoother
is twice as costly under these conditions, but proves more robust against the aspect ratio and becomes the preferred choice for aspect ratios greater than 4.

The proposed multigrid methods present an opportunity to enhance the competitiveness of high-order discontinuous Galerkin methods in more complex applications such as computational fluid dynamics.
Due to its tensor-product structure, the approach offers a straightforward extension to three-dimensional problems, which is the subject to ongoing work.
Further challenges include the development of multigrid preconditioners for variable diffusion and deformed meshes, as demonstrated by \citet{FL04} for the spectral element case.

\section*{Acknowledgements}
Funding by German Research Foundation (DFG) in frame of the project \mbox{STI~157/4-1} is gratefully acknowledged.


\end{document}